%% file: arxiv.tex
\newif\iftr
\newif\ifcnf

\trtrue
\cnffalse

\iftr
  \documentclass[sigplan,screen,nonacm,10pt]{acmart}
  \startPage{1}
  \setcopyright{none}
  \acmConference[]{}{}{}
  \acmISBN{}
  \acmDOI{}
\else
  \documentclass[sigplan,screen,10pt]{acmart}
  \acmConference[Middleware 2021]{}{Dec 6-10, 2021.}{Québec City, Canada}
  \acmYear{2021}
  \acmISBN{} 
  \acmDOI{} 
  \startPage{1}
  \setcopyright{none}
  \makeatletter
  \def\@acmYear{\relax}
  \makeatother

\fi

\AtBeginDocument{%
  \providecommand\BibTeX{{%
    \normalfont B\kern-0.5em{\scshape i\kern-0.25em b}\kern-0.8em\TeX}}}

\usepackage[normalem]{ulem}
\usepackage{soul}

\usepackage[utf8]{inputenc}
\usepackage[T1]{fontenc}
\usepackage{textcomp}
\usepackage{booktabs,tabularx}
\usepackage[binary-units=true]{siunitx}
\usepackage{adjustbox}

\usepackage{listings}
\usepackage{enumitem}
\usepackage{url}

\usepackage{multirow}
\usepackage{booktabs}

\usepackage{fontawesome}
\usepackage{pifont}

\usepackage{tikz}
\definecolor{ao(english)}{rgb}{0.0, 0.5, 0.0}
\newcommand*\circled[1]{\tikz[baseline=(char.base)]{
\node[shape=circle,fill=ao(english),inner sep=1.5pt] (char) {#1};}
}
\DeclareRobustCommand\robustcircled[1]{
  \tikz[baseline=(char.base)]{
    \node[shape=circle,fill=ao(english),inner sep=1.5pt] (char) {#1};
  }
}

\usepackage{subfig}
\usepackage{graphicx}

\usepackage{xargs}
\usepackage{todonotes}

\newcommand\fw{0.31}
\definecolor{darkgreen}{rgb}{0.0, 0.2, 0.13}

\usepackage{makecell}
\usepackage{dblfloatfix}

\usepackage{soul} 
\usepackage{xcolor} 
\soulregister\cite7
\soulregister\autoref7
\soulregister\code7
\soulregister\toolname7
\soulregister\ref7
\soulregister\pageref7
\usepackage{hhline}
\definecolor{lightyellow}{RGB}{250, 250, 180}
\definecolor{HLYELLOW}{RGB}{240, 127, 0}
\definecolor{hlyellow}{RGB}{240, 127, 0}
\sethlcolor{lightyellow}

\begin{document}

\ifcnf
  \title{\vspace{-0.5em}SeBS: A Serverless Benchmark Suite for Function-as-a-Service Computing}
\else
  \title{SeBS: A Serverless Benchmark Suite for Function-as-a-Service Computing}
\fi

\author{Marcin Copik}
\email{marcin.copik@inf.ethz.ch}
\affiliation{%
  \institution{ETH Zürich}
  \country{Switzerland}
}

\author{Grzegorz Kwaśniewski}
\affiliation{%
  \institution{ETH Zürich}
  \country{Switzerland}
}

\author{Maciej Besta}
\affiliation{%
  \institution{ETH Zürich}
  \country{Switzerland}
}

\author{Michał Podstawski}
\affiliation{%
 \institution{ComeOn Group}
 \country{Poland}
}

\author{Torsten Hoefler}
\affiliation{%
  \institution{ETH Zürich}
  \country{Switzerland}
}

\renewcommand{\shortauthors}{Copik, et al.}

\begin{abstract}
\input{secs/abstract}
\end{abstract}

\maketitle

\section{Introduction}
\input{secs/introduction}

\iftr
  \section{Platform Model}
  \label{sec:platform_model}
  \input{secs/faas}

  \section{Benchmark Specification}
  \label{sec:bench_spec}

\input{secs/benchmark-specification}

  \section{Benchmark Implementation}
  \label{sec:bench_impl}

\input{secs/benchmark-implementation}

  \section{Evaluation}
  \input{secs/evaluation}

  \section{Related Work}

\input{secs/related-work}

  \section{Conclusions}
  \input{secs/conclusion}

\else
  \vspace{-0.5em}
  \section{Platform Model}
  \vspace{-0.25em}
  \label{sec:platform_model}

\input{secs/faas}
  \vspace{-0.5em}
  \section{Benchmark Specification}
  \vspace{-0.25em}
  \label{sec:bench_spec}

\input{secs/benchmark-specification}
  \vspace{-0.5em}
  \section{Benchmark Implementation}
  \vspace{-0.25em}
  \label{sec:bench_impl}

\input{secs/benchmark-implementation}
  \vspace{-0.5em}
  \section{Evaluation}
  \vspace{-0.25em}

\input{secs/evaluation}
  \vspace{-0.5em}
  \section{Related Work}
  \vspace{-0.25em}

\input{secs/related-work}
  \section{Conclusions}
  \input{secs/conclusion}
\fi

\begin{acks}
  This project has received funding from the European Research Council (ERC) under the
  European Union's Horizon 2020 programme (grant agreement DAPP, No. 678880),
  and from the German Research Foundation (DFG) and the Swiss National Science Foundation (SNF)
  through the DFG Priority Programme 1648 Software for Exascale Computing (SPPEXA) and the ExtraPeak
  project (Grant Nr. WO1589/8-1).
\end{acks}

\bibliographystyle{ACM-Reference-Format}
\bibliography{cloud,serverless}

\end{document}

%% file: secs/abstract.tex
Function-as-a-Service (FaaS) is one of the most promising directions for the future of cloud services,
and serverless functions have immediately become a new middleware for building scalable
and cost-efficient microservices and applications.
However, the quickly moving technology hinders reproducibility, and
the lack of a standardized benchmarking suite leads to ad-hoc solutions and
microbenchmarks being used in serverless research,
further complicating meta-analysis and comparison of research solutions.
To address this challenge, we propose the Serverless Benchmark Suite:
the first benchmark for FaaS computing that systematically covers
a wide spectrum of cloud resources and applications.  
Our benchmark consists of the specification of representative workloads, the accompanying
implementation and evaluation infrastructure,
and the evaluation methodology that facilitates reproducibility and enables interpretability. 
We demonstrate that the abstract model of a FaaS execution environment ensures the applicability
of our benchmark to multiple commercial providers such as AWS, Azure, and Google Cloud.
Our work facilities experimental evaluation of serverless systems,
and delivers a standardized, reliable and evolving evaluation methodology of performance, 
efficiency, scalability and reliability of middleware FaaS platforms.

%% file: secs/introduction.tex
\ifcnf
  \enlargethispage{\baselineskip}
\fi

Clouds changed the computing landscape with the promise of plentiful
resources, economy of scale for everyone, and on-demand availability without
up-front or long-term commitment.
Reported costs are up to 7$\times$ lower than that of a traditional in-house
server~\cite{Armbrust09abovethe}.
The deployment of middleware in cloud evolved from
a more hardware-oriented \emph{Infrastructure as a Service} (IaaS)
to a more software-oriented \emph{Platform as a Service},
where the cloud service provider takes the responsibility of deploying
and scaling resources~\cite{Binnig:2009:WTT:1594156.1594168}.
\textit{Function-as-a-Service} (FaaS) is a recent development towards
fine-grained computing and billing, where stateless functions are used
to build modular applications without managing infrastructure and
incurring costs for unused services.

\begin{table}
  \footnotesize
  \renewcommand{\arraystretch}{0.5}
	\begin{tabular}{p{3.9cm}p{3.9cm}}
		\toprule
		\multicolumn{2}{c}{\includegraphics[height=1.5\fontcharht\font`\B]{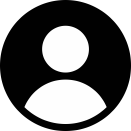}
		  User's perspective} \\
		\midrule
		\faThumbsOUp~ Pay-as-you-go billing & \faThumbsDown~ High 
		computing cost \\
		\faThumbsOUp~ Massive parallelism & \faThumbsDown~ Variable performance \\
		\faThumbsOUp~ Simplified deployment& \faThumbsDown~ Vendor lock-in \\
		\faThumbsOUp~ Architecture agnostic & \faThumbsDown~ Black-box platform\\
		\midrule
		\multicolumn{2}{c}{\includegraphics[height=1.5\fontcharht\font`\B]{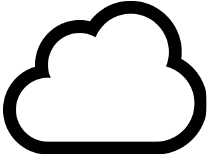}
		 Provider's perspective}\\
		\midrule
    \faThumbsOUp~ Higher machine utilization & \faThumbsDown~ Handling 
		heterogeneity
		\\
		\faThumbsOUp~ Fine-grained scheduling & \faThumbsDown~ 
		Micro-architecture effects  \\
		\bottomrule
  \end{tabular} 
  \caption{\textbf{Intuitive summary of the FaaS model.
  Quantitative measurements are needed for assessing advantages \emph{and}
  disadvantages.}}
	\label{tab:intuitive_comparison}
\ifcnf
  \vspace{-3.5em}
\fi
\end{table}

The flexible FaaS model may be seen as a necessary connection between
ever-increasing demands of diverse workloads on the one hand, and huge data
centers with specialized hardware on the other.
Serverless functions have already become the software glue for building
stateful applications~\cite{10.1145/3361525.3361535,10.14778/3407790.3407836,10.1145/3419111.3421277}.
It is already
adopted by most major commercial providers, such as AWS
Lambda~\cite{awsLambda}, Azure Functions~\cite{azureFunctions}, Google Cloud
Functions~\cite{googleFunctions}, and IBM Cloud Functions~\cite{ibmFunctions},
marking the future of cloud computing.
%
%
From a user perspective, it promises more savings and the pay-as-you-go model where
only active function invocations are billed, whereas a standard IaaS virtual
machine rental incurs costs even when they are idle~\cite{10.5555/3277355.3277369}.
From a provider's perspective,
the fine-grained execution model enables high machine utilization through
efficient scheduling and oversubscription.
Table~\ref{tab:intuitive_comparison} provides an overview of FaaS advantages
and issues.

\begin{figure*}[tbh!]
\ifcnf
  \vspace{-1em}
\fi
  \centering
\includegraphics[width=0.9\textwidth]{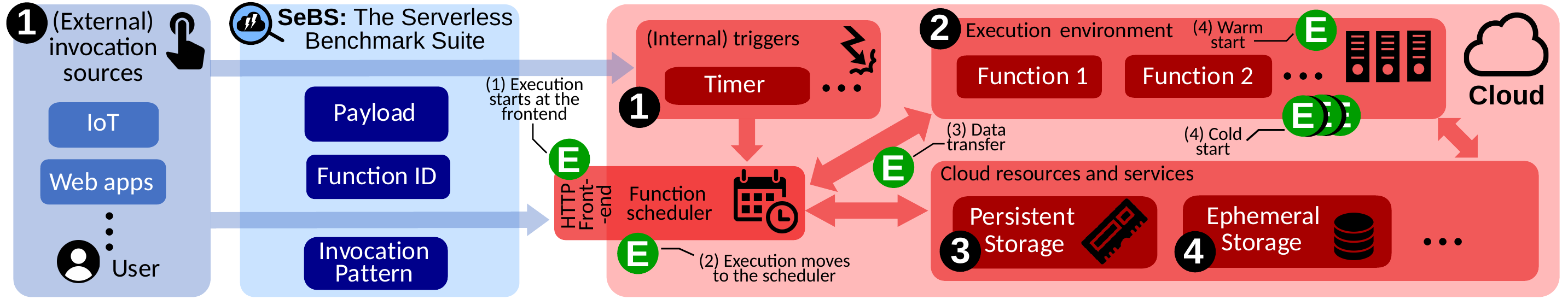}
\ifcnf
  \vspace{-1em}
\fi
\caption{\textbf{An abstract model of a FaaS platform. Labels: \ding{182} - 
triggers, \ding{183} - execution environment, \ding{184} - 
persistent storage, \ding{185} - ephemeral storage,
\robustcircled{\textcolor{white}{E}} - invocation system. Details in Section~\ref{sec:platform_model}.}
}
\ifcnf
  \vspace{-1em}
\fi
 \label{fig:serverless_diagram}
\end{figure*}

%
While serverless computing gained significant traction both in industry and
academia, many authors raised several important issues, for example,
vendor lock-in on commercial platforms, lack of
standardized tools for development and debugging, unpredictable overheads
due to high-latency cold starts~\cite{coldStart}, and surprisingly high
costs of computation-intensive
codes~\cite{DBLP:journals/corr/abs-1902-03383}. Cost and performance
analyses of FaaS applications are further inhibited by the black-box nature of
serverless platforms, and existing analyses rarely generalize beyond a
given vendor or a tested system. Yet, efficient and scalable software
systems cannot be designed without insights into the middleware they're being
built upon.
%
%
%
Hence, there is an urgent need for a benchmarking design that would (1) specify clear
comparison baselines for the evaluation of \emph{many} FaaS workloads on
\emph{different} platforms, (2) enable deriving \emph{general} performance, cost,
and reliability insights about these evaluations, and (3) facilitate the above with
\emph{public and easy to use} implementation.
%

To address these challenges, we introduce the \textbf{Serverless Benchmark Suite (SeBS)}, a FaaS benchmark
suite that combines a systematic cover of a wide spectrum of cloud resources
with detailed insights into black-box serverless platforms.
SeBS comes with a (1) \emph{benchmark specification} based on extensive literature review,
(2) a general \emph{FaaS platform model} for wide applicability,
(3) a set of \emph{metrics} for effective analysis of cost and performance,
(4) \emph{performance models} for generalizing evaluation insights across different cloud infrastructures,
and (5) an \emph{implementation} kit that facilitates evaluating existing
and future FaaS platforms.

We evaluate SeBS on AWS, Microsoft Azure, and Google Cloud Platform. 
Overall, with our benchmarks representing a wide variety of real-world workloads, we provide
the necessary milestone for serverless functions to become an efficient and reliable
software platform for complex and scalable cloud applications.

To summarize, we make the following contributions:

\begin{itemize}[leftmargin=0.5em]

  \item We propose SeBS, a standardized platform for continuous evaluation, analysis, and comparison of FaaS performance, reliability, and cost-effectiveness.

  \item We offer novel \emph{metrics} and \emph{experiments} that, among others,
  enable quantifying the overheads and efficiency of FaaS under various configurations and workloads.

  \item We provide a full \emph{benchmark implementation} and an open-source\footnote{The code is available on GitHub: \texttt{spcl/serverless-benchmarks}.}
\emph{software toolkit} that can automatically build,
deploy, and invoke functions on FaaS systems in AWS, Azure, and GCP, three popular cloud
providers. The toolkit is modular and can be easily extended to
support new FaaS platforms.

\item We provide insights into FaaS performance and consistency (Sec.~\ref{sec:evaluation}, Table~\ref{tab:rw_table_insights}).
We analyze performance and cost overheads of serverless functions, and model
cold start patterns and invocation latencies of FaaS platforms.

\end{itemize}

%% file: secs/faas.tex
We first build a benchmarking model of a FaaS platform that provides an
abstraction of key components, see~\autoref{fig:serverless_diagram}. This enables generalizing design details that
might vary between providers, or that may simply be unknown due to the
black-box nature of a closed-source platform. 

\begin{figure*}[tbh!]
\ifcnf
  \vspace{-1.25em}
	\centering
	\includegraphics[width=1.0\textwidth]{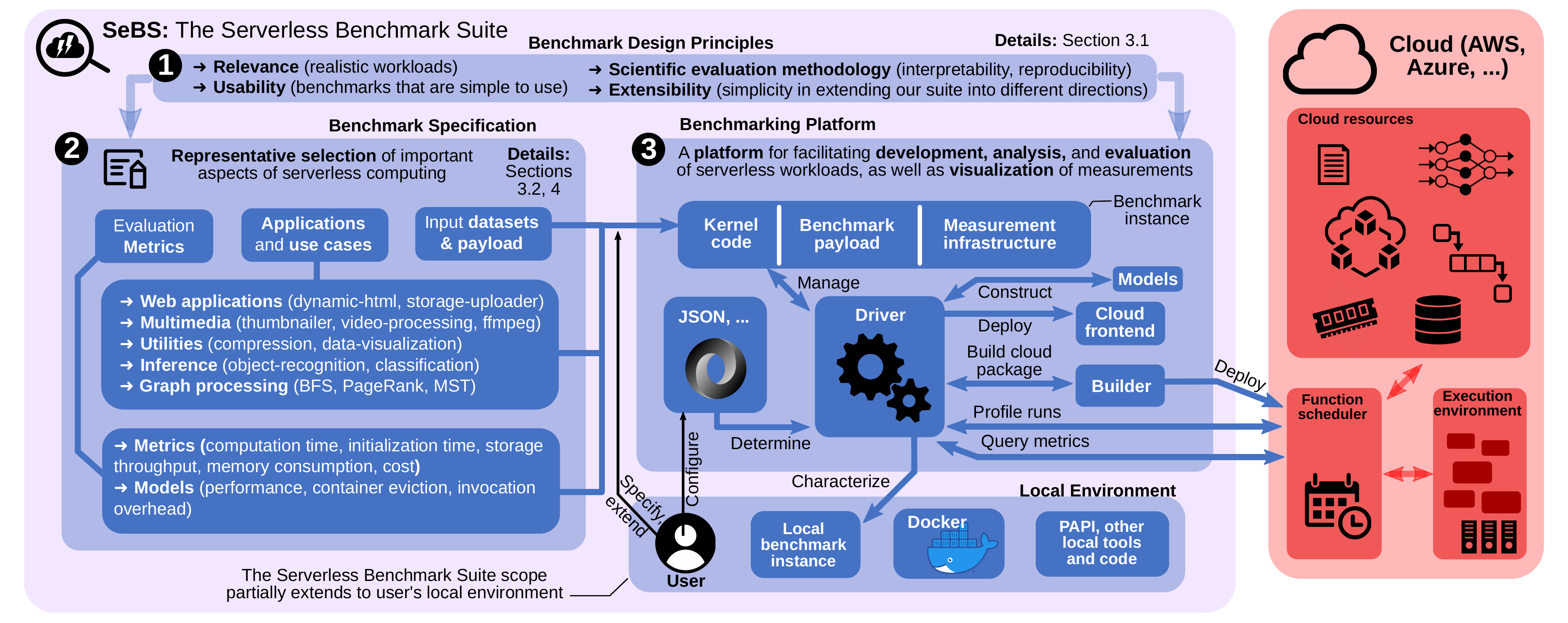}
  \vspace{-2em}
	\caption{\textbf{An overview of the offered serverless benchmark suite.}}
  \vspace{-1em}
	\label{fig:suite_diagram}
\else
	\centering
	\includegraphics[width=1.0\textwidth]{figures/suite_full_slim-less-busy}
	\caption{\textbf{An overview of the offered serverless benchmark suite.}}
	\label{fig:suite_diagram}
\fi
\end{figure*}

\textbf{\ding{182} Triggers. } 
The function lifetime begins with a \emph{trigger}. Cloud providers offer many
triggers to express various ways to incorporate functions into a larger
application. One example is an HTTP trigger, where every request sent to a
function-specific address invokes the function using the passed payload. 
Such triggers are often used for interactive services or
background requests from \emph{Internet-of-Things} (IoT) and edge computing
devices.
Other triggers are invoked periodically, similarly to cron jobs, or
upon events such as a new file upload or a new message in a queue.
Finally, functions
can be triggered as a part of larger FaaS workflows in dedicated services such
as AWS Step and Azure Durable.

\textbf{\ding{183} Execution environment. } 
Functions require a sandbox environment to ensure isolation between
tenants. One option is containers, but
they may incur overheads of up to 20$\times$ over native
execution~\cite{10.1145/3352460.3358296}. Another solution
is lightweight virtual machines (microVMs). They can provide
overheads and bootup times competitive with containers, while improving
isolation and security~\cite{10.1145/3132747.3132763,firecracker}.

\textbf{\ding{184} Persistent Storage. }
The cloud service offers scalable storage and high bandwidth retrieval. 
Storage offers usually consist of multiple containers known as 
\textit{buckets} (AWS, Google) and \textit{containers} (Azure).
Storage services offer high throughput but also high latency for a low 
price,
with fees in the range of a few cents for
\SI{1}{\giga\byte}
%
of data storage, retrieval, or 10,000 write/read operations.

\textbf{\ding{185} Ephemeral Storage. }
The ephemeral storage service addresses the high latency of persistent storage~\cite{216007, 10.5555/3291168.3291200}.
Example use cases include storing
payload passed between consecutive function
invocations~\cite{10.5555/3277355.3277444} and communication in serverless
distributed computing~\cite{DBLP:journals/corr/JonasVSR17}.
The solutions include scalable, in-memory databases offered by the cloud
provider and a custom solution with a VM instance holding an
in-memory, key-value storage.
While the latter is arguably no longer FaaS and the usage of
non-scaling storage services might be considered to be a \textit{serverless
anti-pattern}~\cite{DBLP:journals/corr/abs-1812-03651}, it provides low latency
data storage and exchange platform.

\textbf{\circled{\textcolor{white}{E}} Invocation System. }
The launch process has at least four steps: (a) a
cloud endpoint handling the trigger, (b) the FaaS resource manager and
scheduler deciding where to place the function instance, (c) communication with
the selected cloud server, (d) the server handling invocations with
load-balancing and caching.
A cold start also adds the execution
environment startup latency.
These overheads are hidden from the user but understanding them helps to minimize startup latencies.
Cloud providers benefit from identifying performance bottlenecks in their systems as well.

\ifcnf
  \vspace{-0.5em}
\fi
\section{Serverless Model Analysis}
\label{sec:opportunities}
\ifcnf
  \vspace{-0.25em}
\fi

To design SeBS, we select candidate FaaS workloads and investigate the fundamental limitations that can throttle the migration of some workloads to the serverless environment.

\ifcnf
  \vspace{-0.5em}
\fi
\subsection{Candidate applications}

Workloads with a premise for immediate benefits have 
infrequent invocations, unpredictable and sudden spikes in arriving requests,
and fine-grained parallelism.
%
%
Yet, unprecedented
parallelism offered by FaaS is not simple to harness,
and many workloads struggle to achieve high performance and suffer from problems such as
stragglers~\cite{10.1145/3304112.3325608,10.5555/3154630.3154660,10.1145/3267809.3267815}.
Such FaaS workload classes that may be hard to program for high performance 
are data
analytics~\cite{227653,Mller2019LambadaID,DBLP:journals/corr/abs-1907-11465},
distributed compilation~\cite{234886}, video encoding, linear algebra
and high-performance computing problems~\cite{DBLP:journals/corr/abs-1810-09679}, and machine learning
training and inference~\cite{10.1145/3357223.3362711,8360337,8457817}.

\begin{table*}[t!]\centering
\ifcnf
  \vspace{-1em}
\fi
\footnotesize
\renewcommand{\arraystretch}{0.9}
\begin{tabular}{@{}p{3.0cm}p{4.8cm}p{4.4cm}p{4.5cm}@{}}
	\toprule
\textbf{Policy}  & \textbf{AWS} & \textbf{Azure} & \textbf{GCP} \\
\midrule
\textbf{Languages (native)} & Python, Node.js, C\#, Java, C++, and more. & Python, 
JavaScript, C\#, Java etc. & Node.js, Python, Java, Go\\
\textbf{Time Limit} & 15 minutes & 
10 min / 60 min / Unlimited & 9 minutes\\
\textbf{Memory  Allocation} & Static, 128 - 3008 MB
       & Dynamic, up to 1536 MB & Static, 128, 256, 512, 1024 or 2048 MB\\ 
\multirow{2}{*}{\textbf{CPU Allocation}} & Proportional to memory & 
\multirow{2}{*}{Unknown} & Proportional to memory \\
& 1 vCPU on 1792 MB & & 2.4 GHz CPU at 2048 MB\\

\textbf{Billing} & Duration and declared memory
        & Average memory use, duration
        & Duration, declared CPU and memory \\

\textbf{Deployment}  &  zip package up to 250 MB 
            & zip package, Docker image
            & zip package, up to 100 MB \\

\textbf{Concurrency Limit} & 1000 Functions& 200 Function Apss& 100 Functions\\

\iftr
\textbf{Temporary Disk Space} & 500 MB (must store code package)
                      & Uses Azure Filess
                      & Included in memory usage\\
\fi

\bottomrule
\end{tabular}
\caption{\textbf{Comparison of major commercial FaaS providers - AWS
Lambda~\cite{lambdaLimits}, Azure Functions~\cite{azureLimits} and Google Cloud
Functions~\cite{gcpLimits}. While commercial services have comparable compute
and storage prices, their memory management and billing policies
differ fundamentally.}} \label{tab:aws_azure}
\ifcnf
  \vspace{-2em}
\fi
\end{table*}
\ifcnf
  \vspace{-1.5em}
\fi

\subsection{FaaS model aspects}
\label{sec:challenges}

Although the adaption of serverless computing is increasing in various domains, 
the technical peculiarities that made it popular in the first place are now 
becoming a roadblock for further 
growth~\cite{DBLP:journals/corr/abs-1812-03651,DBLP:journals/corr/abs-1902-03383}.
Both the key advantages and the limitations of FaaS are 
listed in Table~\ref{tab:intuitive_comparison}.
We now describe each aspect to understand the scope of SeBS better.

\textbf{Computing Cost. }
FaaS handles infrequent workloads more cost-effectively than persistent VMs.
Problems such as machine learning training can be much more expensive 
than a VM-based 
solution~\cite{10.5555/3154630.3154660,DBLP:journals/corr/abs-1902-03383},
primarily due to function communication overheads.
FaaS burst parallelism
outperforms virtual machines in data analytics workloads but inflates 
costs~\cite{Mller2019LambadaID}.
Thus, we need to think of  computational performance not only in raw FLOP/s but,
most importantly, as a FLOP/s per dollar ratio.
%
Here, \emph{SeBS includes cost 
efficiency as a primary metric to determine the most efficient configuration 
for a specific workload,
analyze the pricing model's flexibility,
and compare the costs with IaaS approaches.}

\textbf{I/O performance. }
Network I/O affects cold startup latencies,
and it is crucial in ephemeral computing as it relies on external storage.
%
As function instances share the bandwidth on a server machine, the
co-allocation of functions depending on network bandwidth may degrade performance.
Investigation of major cloud providers
revealed significant fluctuations of network and I/O performance,
with the co-location decreasing throughput up to 20$\times$ on AWS~\cite{10.5555/3277355.3277369}.
\emph{SeBS includes network and disk performance as a metric to understand I/O
requirements of serverless functions better.}

\textbf{Vendor Lock-In. }
Lack of standardization in function configuration, deployment, and cloud
services complicates development.
Each cloud provider requires a customization layer that can be non-trivial.
%
%
Tackling this, \emph{SeBS provides a transparent library for 
adapting cloud service interfaces for deployment, invocation, and persistent storage management.}

\textbf{Heterogeneous Environments. }
Major FaaS platforms 
limit user configuration options 
to the amount of memory allocated and an assigned time to access a virtual CPU.
To the best of our knowledge,  
specialized hardware is only offered by nuclio~\cite{nuclio}, a data science-oriented and GPU-accelerated FaaS provider.
While hardware accelerators are becoming key for scalability~\cite{vetter2019extreme},
serverless functions lack an API to allocate and manage such hardware,
similar to solutions in batch systems on HPC clusters~\cite{slurmGRES}.
%
\emph{SeBS includes dedicated tasks that can benefit from specialized 
hardware.}


\textbf{Microarchitectural Hardware Effects. }
The hardware and software stack of server machines is optimized to handle
long-running applications, where major performance challenges include high
pressure on instruction caches or low counts of instructions per cycle
(IPC)~\cite{10.1145/2872887.2750392}. 
The push to microservices lowers the CPU frontend pressure thanks to a smaller code footprint~\cite{10.1145/3297858.3304013}.
Still, they are bounded by single-core performance and frontend inefficiencies
due to high instruction cache miss and branch misprediction rate~\cite{7856643}.
Serverless functions pose new challenges due to a lack of code and data locality.
A microarchitectural analysis of FaaS workloads discovered similar frontend
bottlenecks as in microservices: decreased branch predictor performance and
increased cache misses due to interfering workloads~\cite{10.1145/3352460.3358296}.
%
\textit{SeBS enables low-level characterization of serverless applications
to analyze short-running functions and better understand
requirements for an optimal FaaS execution environment.}

\ifcnf
  \vspace{-1.5em}
\fi
\subsection{FaaS platforms' limitations}
\label{sec:faas_platforms}


To support the new execution model, cloud providers put restrictions on user's
code and resource consumption.
%
%
Although some of those restrictions can be
overcome, developers must design
applications with these
limitations in mind.
Table~\ref{tab:aws_azure} presents a detailed overview of three commercial Faas platforms:
AWS Lambda~\cite{awsLambda},
Azure Functions~\cite{azureFunctions},
and Google Cloud Functions~\cite{googleFunctions}.
%
%
Azure Functions change the semantics with an introduction
of \emph{function apps} that consists of multiple functions.
The functions are bundled and deployed together, and a single function app instance
can use processes and threads to handle multiple function instances from the same app.
Thus, they benefit from less frequent cold starts and increased
locality while not interfering with isolation and security requirements.

%% file: secs/benchmark-specification.tex
We first discuss principles of reliable benchmarking used in SeBS (Section~\ref{sec:benchmarking-cloud}).
Then, we propose SeBS' specification by analyzing the scope of common FaaS workloads and classifying
them into six major categories (Section~\ref{sec:applications}).

\subsection{Benchmark Design Principles}
\label{sec:benchmarking-cloud}


Designing benchmarks is a difficult ``dark
art''~\cite{10.1007/978-3-642-36727-4_12}. For SeBS, we follow well-known guidelines~\cite{v.Kistowski:2015:BB:2668930.2688819,Binnig:2009:WTT:1594156.1594168,benchmarking}.

%
\textbf{Relevance. }
%
%
%
%
%
%
%
We carefully inspect serverless 
use cases in the literature to select representative workloads that
stress different components of a FaaS platform. We focus on core FaaS
components that are widely used on all platforms, and are
expected to stay relevant for the foreseeable future.

%
\textbf{Usability. }
Benchmarks that are easy to run benefit from a high degree of self-validation~\cite{v.Kistowski:2015:BB:2668930.2688819}.
In addition to a benchmark specification, we provide \emph{a benchmarking platform} and a
\emph{reference implementation} to enable automatic deployment and performance
evaluation of cloud systems, minimizing the configuration and preparation
effort from the user. 
%

%
%

%
\textbf{Reproducibility \& Interpretability.} 
%
%
For reproducibility and interpretability of outcomes,
we follow established guidelines for scientific benchmarking of
parallel codes~\mbox{\cite{hoefler2015scientific}}.
We compute the 95\% and 99\% non-parametric confidence intervals~\mbox{\cite{hoefler2015scientific,10.5555/1996385}}
and choose the number of samples such that intervals are within 5\% of the median.
Still, in multi-tenant systems with shared infrastructure,
one cannot exactly reproduce the system state and achieve performance.
The FaaS paradigm introduces further challenges
with a lack of control on function placement. 
Thus, in SeBS,
we also focus on understanding and minimizing the deviations of measured
values. 
For example, we consider the geolocation of cloud resources and the
time of day when running experiments. This enables us to minimize effects such
as localized spikes of a cloud activity when many users use it.

%
\textbf{Extensibility.} 
While the SeBS implementation uses existing cloud services and relies on
interfaces specific to providers, the specification of SeBS depends
only on the abstract FaaS model from Section~\ref{sec:platform_model}.
Thus, we do not lock the benchmark in a dependency on a specific commercial
system. 

\subsection{Applications}
\label{sec:applications}

\begin{table}\centering
\footnotesize
\setlength{\tabcolsep}{1.5pt}
\begin{tabular}{@{}p{1.5cm}p{4.5cm}lll@{}}\toprule
  \textbf{Type} & \textbf{Name} & \textbf{Language} & 
  \textbf{Deps} \\
\cline{1-5}
  \multirow{4}{*}{Webapps} & \multirow{2}{*}{dynamic-html} &
    Python & jinja2 \\
    & & Node.js & mustache \\
\cline{3-5}
    & \multirow{2}{*}{uploader} & Python & - \\
    & & Node.js & request \\
\cline{1-5}

  \multirow{3}{*}{Multimedia} & \multirow{2}{*}{thumbnailer} & 
    Python & Pillow \\
    & & Node.js & sharp \\
  \cline{3-5}
    & video-processing & Python & \textbf{ffmpeg} \\
\cline{1-5}
    \multirow{2}{*}{Utilities} & compression & Python & - \\
    & data-vis & Python & squiggle \\
    \cline{3-5}
    \cline{1-5}
  \multirow{1}{*}{Inference} & \multirow{1}{*}{image-recognition} & 
    Python & pytorch \\
\cline{1-5}
    \multirow{3}{*}{Scientific} & \multirow{1}{*}{graph-pagerank} &
    \multirow{3}{*}{Python} & \multirow{3}{*}{igraph} \\
    & \multirow{1}{*}{graph-mst} &&& \\
    & \multirow{1}{*}{graph-bfs} &&& \\
    \bottomrule
\end{tabular}
\caption{\textbf{SeBS applications. One application - \emph{video.processing} - 
requires a non-pip package: ffmpeg  (marked in bold).}}
\label{tab:applications}
\ifcnf
  \vspace{-3em}
\fi
\end{table}

Our collection of serverless applications is in~\autoref{tab:applications}.
They represent different performance profiles, from simple website backends
with minimal CPU overhead to compute-intensive machine learning
tasks. To accurately characterize each application's requirements, we conduct
a local, non-cloud evaluation of application metrics describing requirements on
computing, memory, and external resources (Section~\ref{sec:bench_impl}). The
evaluation allows us to classify applications, verify that our benchmark set is
representative, and pick benchmarks according to the required resource
consumption.


%
\textbf{Web Applications. } 
FaaS platforms allow building simplified static websites
where dynamic features can be offloaded to a serverless backend. We include two
examples of small but frequently involved functions: \emph{dynamic-html} (dynamic HTML generation from a
predefined template) and
\emph{storage-uploader} (upload of a
file from a given URL to cloud storage).
%
%
They
have low requirements on both CPU and memory.

%
\textbf{Multimedia. } 
A common serverless workload is processing multimedia
data. Images uploaded require the creation of thumbnails, as we do in our
benchmark kernel \emph{thumbnailer}.
Videos are usually processed to compress,
extract audio, or convert to more suitable formats. We include an application \emph{video-processing}
that uses a static build of \emph{ffmpeg} to apply a watermark to a video and convert
it to a gif file.

%
\textbf{Utilities. } 
Functions are used as backend processing tools for too complex problems
for a web server or application frontend. We consider \emph{compression} and \emph{data-vis}.
In the former, the function compresses a set of files and returns an archive to the user,
as seen in online document office suites and text editors.
We use acmart-master template as evaluation input.
In the latter, we include the backend of DNAVisualization.org~\cite{dna_vis, 10.1093/nar/gkz404},
an open-source website providing serverless visualization of DNA sequences, using the squiggle Python library~\cite{Lee2018}. The website
passes DNA data to a function which generates specified visualization and caches results in the storage.

%
\textbf{Inference. }
Serverless functions implement machine learning inference
tasks for edge IoT devices and websites to handle scenarios such as image processing
with object recognition and classification. We use as an example a standard image recognition
with pretrained ResNet-50 model served with the help of pytorch~\cite{NEURIPS2019_9015} and,
for evaluation, images from \textit{fake-resnet} test from MLPerf inference benchmark~\cite{reddi2019mlperf}.
Deployment of PyTorch requires additional steps to ensure that the final
deployment package meets the limits on the size of the code package. In our case, the
most strict requirements are found on AWS Lambda with a limit of
%
250 megabytes
of uncompressed code size.
We fix the PyTorch version to 1.0.1 with torchvision in version 0.3.
We disable all accelerator support (only CPU), strip shared libraries, and remove tests and binaries from the package.
While deep learning frameworks can provide lower inference latency with GPU processing,
dedicated accelerators are not currently widely available on FaaS platforms,
as discussed in Section~\ref{sec:challenges}.

%
\textbf{Scientific.}
As an example of scientific workloads, we consider irregular graph
computations, a more recent yet established class of
workloads~\cite{lumsdaine2007challenges, besta2017push, sakr2020future,
besta2019substream}. We selected three important problems: Breadth-First
Search (BFS)~\cite{beamer2013direction, besta2017slimsell}, PageRank
(PR)~\cite{page1999pagerank}, and
%
%
Minimum Spanning Tree (MST).
BFS is used in many more complex schemes (e.g., 
in computing maximum flows~\cite{ford2009maximal}), it represents a large
family of graph traversal problems~\cite{besta2017slimsell}, and it is a basis of the
Graph500 benchmark~\cite{murphy2010introducing}.
PR is a leading scheme for ranking websites and it stands for a class of
centrality problems~\cite{brandes2007centrality, solomonik2017scaling}.
%
%
MST is used in many analytics and engineering problems, and represents graph
optimization problems~\cite{papadimitriou1998combinatorial,
gianinazzi2018communication, besta2020high}.
All three have been extensively research in a past
decade~\cite{beamer2013direction, grygorash2006minimum, besta2019slim,
besta2018log, schweizer2015evaluating, besta2015accelerating,
besta2019demystifying, berkhin2005survey}.
We select the corresponding algorithms such that
they are all data-intensive but differ in the details of the workload
characteristics (e.g., BFS, unlike PR, may come with severe work imbalance across iterations).
%

%% file: secs/benchmark-implementation.tex
We complement the benchmark specification introduced in the previous section with
our benchmarking toolkit.
We discuss the set of metrics used to characterize application requirements
and measure performance overheads (Section~\ref{sec:metrics}).
SeBS enables automatic deployment and invocation of benchmarks,
specified in the previous section
(Section~\ref{sec:implementation}).
This benchmarking platform is used for parallel experiments
that model and analyze the behavior of FaaS systems (Section~\ref{sec:evaluation}).

\subsection{Application Metrics}
\label{sec:metrics}

We now discuss in detail metrics that are measured locally and in the cloud execution.

\noindent
\textit{Local metrics. } These metrics provide an accurate 
profile of application performance and resource usage to the 
user.

\begin{itemize}[leftmargin=0.5em]
\item \textbf{Time.} We measure execution time to find which applications
require significant computational effort, and we use hardware performance
counters to count instructions executed, a metric less likely influenced by
system noise.

\item \textbf{CPU utilization.} 
We measure the ratio of time spent by the application on the CPU, both in the
user and the kernel space, to the wall-clock time. This metric helps to detect
applications stalled on external resources.

\item \textbf{Memory.} 
Peak memory usage is crucial for determining application configuration and
billing.
It also enables providers 
to bound the number of active or
suspended containers.
Instead of resident set size (RSS) which overapproximates actual
memory consumption, we measure the unique set size (USS) and
proportional set size (PSS).
Thus, we enable an analysis of benefits from page sharing.

\item \textbf{I/O.} 
I/O intensive functions  may be affected by contention.
Average throughput of filesystem I/O and network
operations decreases with the number of co-located function invocations that
have to share the bandwidth, leading to significant network performance
variations~\cite{10.5555/3277355.3277369}.

\item \textbf{Code size.} 
The size and complexity of dependencies impact the warm \emph{and}
cold start latency. Larger code packages increase deployment time from cloud
storage and the warm-up time of language runtime.
\end{itemize}
%
\textit{Cloud metrics. }
The set of metrics available in the cloud is limited because of the black-box
nature of the FaaS system.
Still, we can gain additional information through microbenchmarks and modeling experiments (Section~\ref{sec:evaluation}).
\begin{itemize}[leftmargin=0.5em]
\item \textbf{Benchmark, Provider and Client Time.} 
We measure execution time on three levels: directly measure benchmark execution time in cloud,
including work performed by function, but not network and system latencies; query cloud provider
measurements, adding overheads of language and serverless sandbox; measure end-to-end execution
latency on client side, estimating complete overhead with the latency of function scheduling and
deployment.


\item \textbf{Memory.} 
The actual memory consumption plays a crucial role in determining cost
on platforms with dynamic memory allocation. Elsewhere, the peak memory consumption
determines the execution settings and billing policies.

\item \textbf{Cost.} 
The incurred costs are modeled from billed duration, memory consumption,
and a number of requests made to persistent storage. While AWS  
enables estimating the cost of each function execution, Azure offers a 
monitoring service with query interval not shorter than one second.
\end{itemize}

\subsection{Implementation}
\label{sec:implementation}

We implement the platform from Figure~\ref{fig:suite_diagram} to fulfill three
major requirements: application characterization, deployment to the cloud, and
modeling of cloud performance and overheads. We describe SeBS modularity
and the support for the inclusion of new benchmarks, metrics, and platforms.

\textbf{Deployment.}
SeBS handles all necessary steps of invoking a function in the cloud.
We allocate all necessary resources and do not use third-party dependencies,
such as the Serverless framework~\cite{serverless}, since a flexible and
fine-grained control over resources and functions is necessary to efficiently handle
large-scale and parallel experiments, e.g., the container eviction model (Sec.~\ref{sec:evaluation}).
For each platform, we implement the simplified interface described below.
Furthermore, benchmarks and their dependencies are built within Docker containers resembling
function execution workers to ensure binary compatibility with the cloud.
Google Cloud Functions use the cloud provider Docker-based build system as required by the provider.
\emph{SeBS can be extended with new FaaS platforms by implementing the described interface
and specifying Docker builder images.}
%
\begin{lstlisting}[language=Python]
class FaaS:
 def package_code(directory, language: [Py, JS])
 def create_function(fname, code, lang: [Py, JS]), config)
 def update_function(fname, code, config)
 def create_trigger(fname, type: [SDK, HTTP])
 def query_logs(fname, type: [TIME, MEM, COST])
\end{lstlisting}

\textbf{Benchmarks.}
We use a single benchmark implementation in a high-level language for all cloud providers.
Each benchmark includes a Python function to generate inputs for invocations of
varying sizes.
SeBS implements provider-specific wrappers for entry functions to support different input formats and interfaces.
Each benchmark can add custom build actions, including installation of native dependencies
and supporting benchmark languages with a custom build process,
such as the AWS Lambda C++ Runtime.
\emph{New applications integrate easily into SeBS:
the user specifies input generation procedure, configures dependencies and optional build actions,
and adjusts storage access functionalities.}
%
\begin{lstlisting}[language=Python]
def function_wrapper(provider_input, provider_env)
 input = json(provider_input)
 start_timer()
 res = function()
 time = end_timer()
 return json(time, statistics(provider_env), res)
\end{lstlisting}

%
\textbf{Storage.}
We use light-weight wrappers to handle different storage APIs used by cloud providers.
Benchmarks use the SeBS abstract storage interface, and we implement one-to-one
mappings between our and provider's interface.
The overhead is limited to a single redirect of a function call.
\emph{New storage solutions require implementing a single interface, and benchmarks will use it automatically.}

\textbf{Experiments}
SeBS implements a set of experiments using provided FaaS primitives.
Experiments invoke functions through an abstract trigger interface, and we implement
cloud SDK and HTTP triggers.
The invocation result includes SeBS measurements and an unchanged output of the benchmark application.
SeBS metrics are implemented in function wrappers and with the provider log querying facilities.
Each experiment includes a postprocessing step that examines execution results and provider logs.
\emph{New experiments and triggers are integrated automatically into SeBS through
  a common interface. SeBS can be extended with new types of metrics by plugging measurement code
  in SeBS benchmark wrappers, by using the provided log querying facilities,
and by returning benchmark-specific measurements directly from the function.}

\textbf{Technicalities.}
We use Docker containers with language workers in Python and Node.js in local evaluation;
minio~\cite{minio} implements persistent storage.
We use PAPI~\cite{10.1007/978-3-642-11261-4_11} to gather
low-level characteristics (we found the results from Linux \emph{perf} to be unreliable when the application lifetime is short).
For cloud metrics, we use provider's API to query execution time, billing, and memory consumption, when available.
We use \emph{cURL} to exclude the HTTP connection overheads for client time measurements. 
We enforce cold starts by updating function configuration on AWS and by
publishing a new function version on Azure and GCP.

%% file: secs/evaluation.tex
\begin{figure*}
\ifcnf
  \vspace{-1.25em}
  \includegraphics[width=\textwidth]{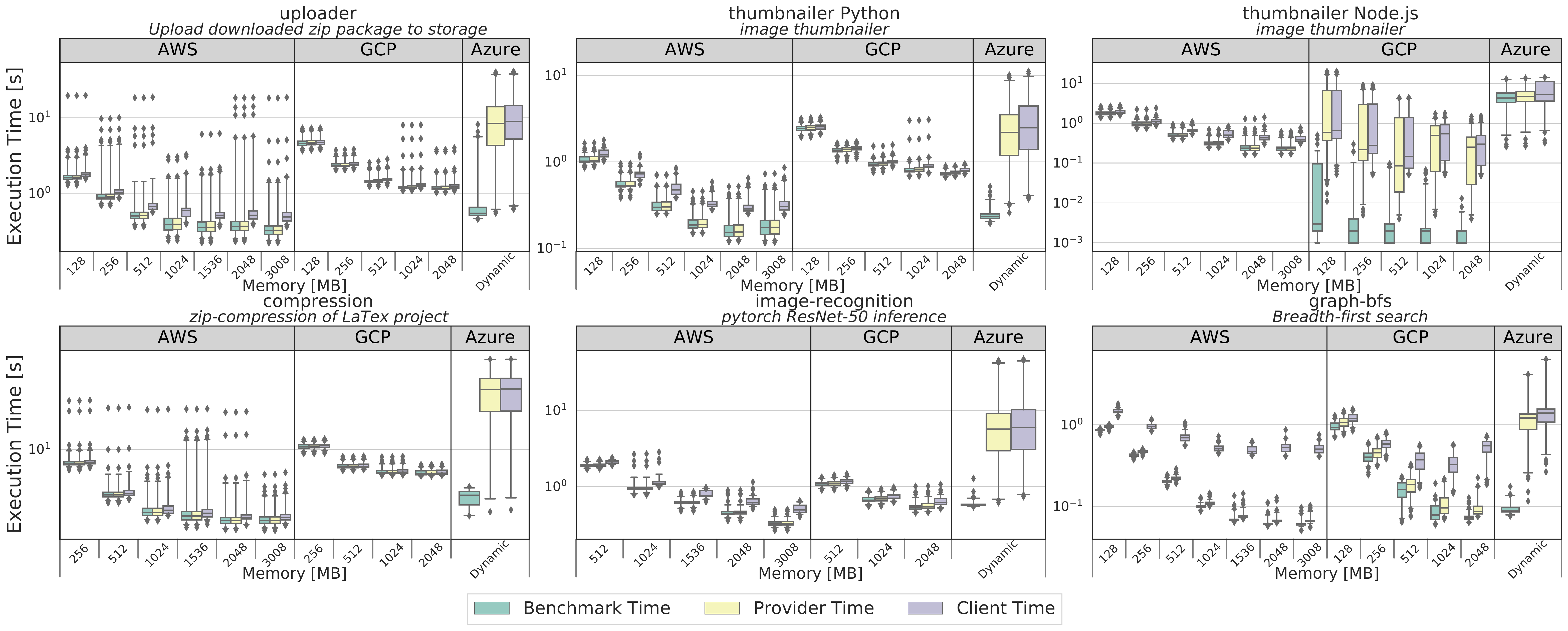}
  \vspace{-2.5em}
    \caption{\textbf{Performance of SeBS applications on AWS Lambda, 
    Azure Functions and Google Cloud Functions. Experiment includes 200 \emph{warm} invocations.
    Whiskers include data from 2th to 98th percentile.}}
	\label{fig:time_evaluation}
  \vspace{-1em}
\else
  \includegraphics[width=\textwidth]{figures/time.pdf}
    \caption{\textbf{Performance of SeBS applications on AWS Lambda, 
    Azure Functions and Google Cloud Functions. Experiment includes 200 \emph{warm} invocations.
    Whiskers include data from 2th to 98th percentile.}}
	\label{fig:time_evaluation}
\fi
\end{figure*}

\label{sec:evaluation}

We show how SeBS provides a consistent and accurate methodology of comparing serverless
providers, and assessing the FaaS performance, reliability, and applicability to various classes of workloads.
We begin with a local benchmark evaluation to verify that we cover different computing
requirements (Section~\ref{sec:benchmarkcharacteristics}).
Next, we thoroughly evaluate serverless systems' performance-cost tradeoffs (Section~\ref{sec:performance},~\ref{sec:cost}).
We determine the platforms with the best and most consistent performance and analyze
serverless computing's suitability for different types of workloads. 
Finally, we use SeBS to better understand serverless platforms through performance
modeling of invocation latencies (Section~\ref{sec:invocation}), and container eviction
policies (Section~\ref{sec:eviction}).
two major black-box components affecting FaaS applicability as middleware for reliable and
scalable applications. We summarize new results and insights provided by SeBS in Table~\ref{tab:rw_table_insights}.

\textbf{Configuration } 
We evaluate SeBS on the three most representative FaaS platforms:
the default AWS Lambda plan without provisioned concurrency,
the standard Linux consumption plan on Azure Functions, and Google Cloud Functions,
in regions \emph{us-east-1}, \emph{WestEurope}, and \emph{europe-west1}, respectively.
We use S3, Azure Blob Storage, and Google Cloud Storage for persistent storage,
HTTP endpoints as function triggers,
and deploy Python 3.7 and Node.js 10 benchmarks.

%
\subsection{Benchmark Characteristics}
\label{sec:benchmarkcharacteristics}

\begin{table}\centering
\begin{adjustbox}{max width=\linewidth}
\begin{tabular}{lllllll@{}}\toprule
    Name & Lang. & Cold Time [ms] & Warm Time [ms] & Instructions& CPU\% \\
\cline{1-6}
    \multirow{2}{*}{dynamic-html} &
    \textbf{P} &
    $130.4 \pm 0.7$ & $1.19 \pm 0.01$  & $7.02M \pm 287K$ & 99.4\% \\
    & \textbf{N} &
    $84 \pm 2.8$ & $0.28 \pm 0.5$ & - & 97.4\% \\
\cline{1-2}
     \multirow{2}{*}{uploader} & \textbf{P} &
    $236.9 \pm 12.7$ & $126.6 \pm 8.9$ & $94.7M \pm 4.45M$ & 34\% \\
       & \textbf{N} & 
    $382.8 \pm 8.9$ & $135.3 \pm 9.6$ & - & 41.7\%\\
\cline{1-2}

  \multirow{2}{*}{thumbnailer} & 
  \textbf{P} &
    $205 \pm 1.4$ & $65 \pm 0.8$ & $404M \pm 293K$ & 97\% \\
                  & \textbf{N} &
    $313 \pm 4$ & $124.5 \pm 4.4$ &  - & 98.5\% \\
\cline{1-2}
    video-processing & \textbf{P} & 
    $1596 \pm 4.6$ & $1484 \pm 5.2$ &  - &  - \\
    compression & \textbf{P} &

    $607 \pm 5.3 $ & $470.5 \pm 2.8$ & $1735M \pm 386K$ &  88.4\% \\
image-recognition & \textbf{P} & $1268 \pm 74$ & $124.8 \pm 2.7$ & $621M \pm 278K$ & 98.7\% \\
    \multirow{1}{*}{graph-pagerank} &
    \multirow{3}{*}{\textbf{P}} & $194 \pm 0.8$ & $106 \pm 0.3$ & $794M \pm 293K $& 99\% \\
    \multirow{1}{*}{graph-mst} && $125 \pm 0.8$ & $38 \pm 0.4$ & $234M \pm 289K$& 99\% \\
    \multirow{1}{*}{graph-bfs} && $123 \pm 1.1$ & $36.5 \pm 0.5$& $222M \pm 300K$ & 99\% \\
    \bottomrule
\end{tabular}
\end{adjustbox}
\caption{Standard characterization of \textbf{P}ython and \textbf{N}ode.js benchmarks over 50 executions in a local environment on AWS \emph{z1d.metal} machine.}
\label{tab:applications_evaluation}
\ifcnf
\vspace{-3em}
\fi
\end{table}
We begin with a local evaluation summarized in~\mbox{\autoref{tab:applications_evaluation}}.
We selected applications representing different performance profiles,
from website backends with minimal CPU overhead and up to compute-intensive
machine learning inference.
The evaluation allows us to classify
applications, verify that our benchmark set is representative and select to experiments benchmarks accordingly
to required resource consumption.

\begin{figure*}
    \includegraphics[width=\textwidth]{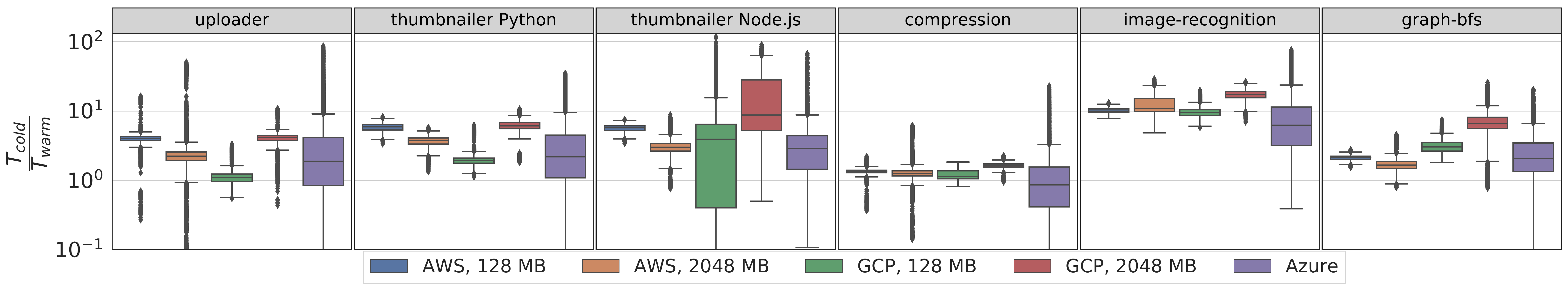}
\ifcnf
    \vspace{-2em}
\fi
    \caption{\textbf{Cold startup overheads of benchmarks on AWS Lambda and Google Cloud
    Functions, based on \emph{cold} and \emph{warm} executions (Figure~\ref{fig:time_evaluation}).}
    }
	\label{fig:cold_startup}
\ifcnf
  \vspace{-1em}
\fi
\end{figure*}

\ifcnf
\vspace{-0.5em}
\fi
\subsection{Performance analysis}
\label{sec:performance}
\ifcnf
\vspace{-0.25em}
\fi


We design a benchmarking experiment \emph{Perf-Cost} to measure the cost and performance of FaaS executions.
We run concurrent function invocations, sampling to obtain \emph{N}
\emph{cold} invocations by enforcing container eviction between each invocations batch.
Next, we sample the function executions to obtain N calls to a \emph{warm} container.
We measure client, function, and provider time (Section~\ref{sec:metrics}).
We compute non-parametric confidence intervals\mbox{~\cite{10.1145/2807591.2807644}} for client time
and select the number of samples \textit{N} = 200 to ensure that intervals are within 5\% of the median for AWS
while the experiment cost stays negligible.
We perform 50 invocations in each batch to include invocations in different sandboxes,
and use the same configuration
on Azure and GCP for a fair and unbiased comparison of performance
and variability.
\footnote{We generate more samples due to unreliable cloud logging services.
We always consider first 200 correctly generated samples and don't skip outliers.}

We benchmark network bandwidth (\emph{uploader}),
storage access times and compute performance (\emph{thumbnailer}, \emph{compression}),
large cold start deployment and high-memory compute (\emph{image-recognition}),
significant output returned (\emph{graph-bfs}),
and compare performance across languages (Python and Node.js versions of \emph{thumbnailer}).

%
%
\textbf{\emph{\textbf{Q1} How serverless applications perform on FaaS platforms?}}
Figure~\mbox{\ref{fig:time_evaluation}} presents significant differences in warm invocations
between providers, with AWS Lambda providing the best performance on all benchmarks.
Each function's execution time decreases until
it reaches a plateau associated with sufficient resources to achieve highest observable
performance.
Only benchmark \emph{graph-bfs} achieves comparable performance on the Google platform,
with the largest slowdown observed on benchmarks relying on storage bandwidth (\emph{thumbnailer}, \emph{compression}).
%
On Azure, we note a significant difference between benchmark and provider times
on Python benchmarks.
To double-check our measurements' correctness and verify
if initialization overhead is the source of such discrepancy, we sequentially repeat warm invocations
instead of using concurrent benchmark executions.
The second batch presents more stable measurements, and we observe performance comparable
to AWS on computing benchmarks \emph{image-recognition} and \emph{graph-bfs}.

Our results verify previous findings that CPU and I/O allocation increases with the memory
allocation~\mbox{\cite{10.5555/3277355.3277369}}.
However, our I/O-bound benchmarks (\emph{uploader}, \emph{compression})
  reveal that the distribution of latencies is much wider and includes many outliers,
  which prevents such functions from achieving consistent and predictable performance.

\emph{
  Conclusions: AWS functions consistently achieve the highest performance.
  Serverless benefits from larger resource allocation, but selecting the right configuration
  requires an accurate methodology for measuring short-running functions.
  I/O-bound workloads are not a great fit for serverless.
}

%
%
\emph{\textbf{Q2 How cold starts affect the performance?} }
%
%
We estimate cold startup overheads by considering all $N^2$ combinations of $N$ cold and
$N$ warm measurements.
%
In Figure~\mbox{\ref{fig:cold_startup}}, we summarize
the ratios of cold and warm client times of each combination.
This approach doesn't provide a representative usage of Azure Functions,
where a single function app instance handles multiple invocations.
To estimate real-world cold startups there, instead of \emph{cold} runs, we use
concurrent \emph{burst} invocations that include cold and warm executions.

We notice the largest cold startup overheads on benchmark \emph{image-recognition} (large deployment,
download model from storage), where a cold
execution takes on average up to ten times longer than a warm invocation, which correlates with
previous findings (c.f.~\cite{Manner2018ColdSI}).
Simultaneously, the \emph{compression} benchmark shows that cold start can have a negligible impact
for longer running functions (> 10 seconds).
Azure provides lower overheads, with the highest gains on benchmarks with a large deployment
package and long cold initialization, at the cost of higher variance.

However, we notice an unusual and previously not reported contrast between Amazon and Google platforms:
while high memory invocations help to mitigate cold startup overheads on Lambda, providing more
CPU allocation for initialization and compilation, they have an adverse
effect on Google Functions, except for benchmark \emph{image-recognition} discussed above.
A possible explanation of this unexpected result might be a smaller pool
of more powerful containers, leading to higher competition between invocations and longer allocation
times.
\emph{Conclusions: more powerful (and expensive) serverless allocations are not a generic
  and portable solution to cold startup overheads. Functions with expensive cold
initialization benefit from functions apps on Azure.}

%
\emph{\textbf{Q3 FaaS performance: consistent and portable?} }
%
%
Vendor lock-in is a major problem in serverless.
We look beyond the usual concern of provider-specific services,
and examine changes in function's performance and availability.

\textbf{Performance deviations}
%
In Figure~\ref{fig:time_evaluation}, we observe
the highest variance in benchmarks relying on I/O bandwidth (\emph{uploader} and
\emph{compression}).
Compute-intensive applications show consistent execution times (\emph{image-recognition})
while producing a notable number of stragglers on long-running functions (\emph{compression}).
Function runtime is not the primary source of variation
since we don't observe significant performance differences between Python and Node.js.
Google's functions produced fewer outliers on warm invocations.
Contrarily, Azure's results present significant performance deviations.
Provider and client time measurements were significantly higher and more variant
than function time on all benchmarks, except Node.js one,
implying that the Python function app generates observed variations.
The Node.js benchmark shows a very variable performance,
indicating that invocations might be co-located in the
same language worker.
Finally, we consider the network as a source of variations.
The ping latencies to virtual machines allocated in the same resource region as benchmark functions
were consistent and equal to 109, 20, and 33 ms on AWS, Azure, and GCP, respectively.
Thus, the difference between client and provider times
cannot be explained by network latency only.

\textbf{Consistency}
On AWS, consecutive warm invocations always hit warm containers, even when the number of concurrent calls is large.
On the other hand, GCP functions revealed many unexpected cold startups, even if consecutive calls never overlap.
The number of active containers can increase up to 100 when processing batches of 50 requests.
Possible explanations include slower resources deallocation and a delay in notifying
the scheduler about free containers.

\textbf{Availability}
Concurrent invocations can fail due to service unavailability, as observed occasionally on Azure and Google Cloud.
On the latter, \emph{image-recognition} generated up to 80\% error rate on 4096 MB memory when processing 50
invocations, indicating a possible problem with not sufficient cloud resources to process our requests.
Similarly, our experiments revealed severe performance degradation on Azure when handling
concurrent invocations, as noted in Section~\ref{sec:performance}.Q1, with long-running
benchmark \emph{compression} being particularly affected.
While Azure can deliver an equivalent performance for sequential invocations,
it bottlenecks on concurrent invocations of Python functions.

\textbf{Reliability}
GCP functions occasionally failed due to exceeding the memory limit,
as was the case for benchmarks \emph{image-recognition} and \emph{compression} on 512 MB and 256 MB, respectively.
Memory-related failure frequency was 4\% and 5.2\%, and warm invocations of
\emph{compression} had recorded 95th and 99th percentile of memory consumption
as 261 and 273 MB, respectively.
We didn't observe any issues with the same benchmarks and workload on AWS,
where the cloud estimated memory consumption as a maximum of 179 MB and exactly 512 MB, respectively.
While the memory allocation techniques could be more lenient on AWS, the GCP
function environment might not free resources efficiently.

\emph{Conclusions: the performance of serverless functions is not stable, and an identical software
  configuration does not guarantee portability between FaaS providers. GCP users
suffer from much more frequent reliability and availability issues.
}

\begin{table}\centering
	\begin{adjustbox}{max width=\linewidth}
  \begin{tabular}{lllllll}\toprule
    & \textbf{Upl} & \textbf{Th, Py} & \textbf{Th, JS} & \textbf{Comp} & \textbf{Img-Rec} & \textbf{BFS} \\
		\midrule
      IaaS, Local [s] & 0.216 & 0.045 & 0.166 & 0.808 & 0.203 & 0.03\\
      IaaS, S3 [s] & 0.316 & 0.13 & 0.191 & 2.803 & 0.235 & 0.03\\
      FaaS [s] & 0.389 & 0.188 & 0.253 & 2.949 & 0.321 & 0.075 \\
      Overhead & 1.79x & 4.14x & 1.43x & 3.65x & 1.58x & 2.49x\\
      Overhead, S3 & 1.23x & 1.43x & 1.24x & 1.05x & 1.37x & 2.4x\\
      Mem [MB] & 1024 & 1024 & 2048 & 1024 & 3008 & 1536\\
		\bottomrule
	\end{tabular}
\end{adjustbox}
	\caption{\textbf{Benchmarks performance on AWS Lambda and AWS EC2 t2.micro instance. Median from 200 warm executions.}}
	\label{tab:iaas_performance}
\ifcnf
  \vspace{-2.5em}
\fi
\end{table}

\begin{figure*}
	\centering
	\resizebox*{0.95\width}{0.92\totalheight}{
		\subfloat[Compute cost of 1M invocations (USD).]{%
      \includegraphics[width=\dimexpr0.66\textwidth,height=5cm]{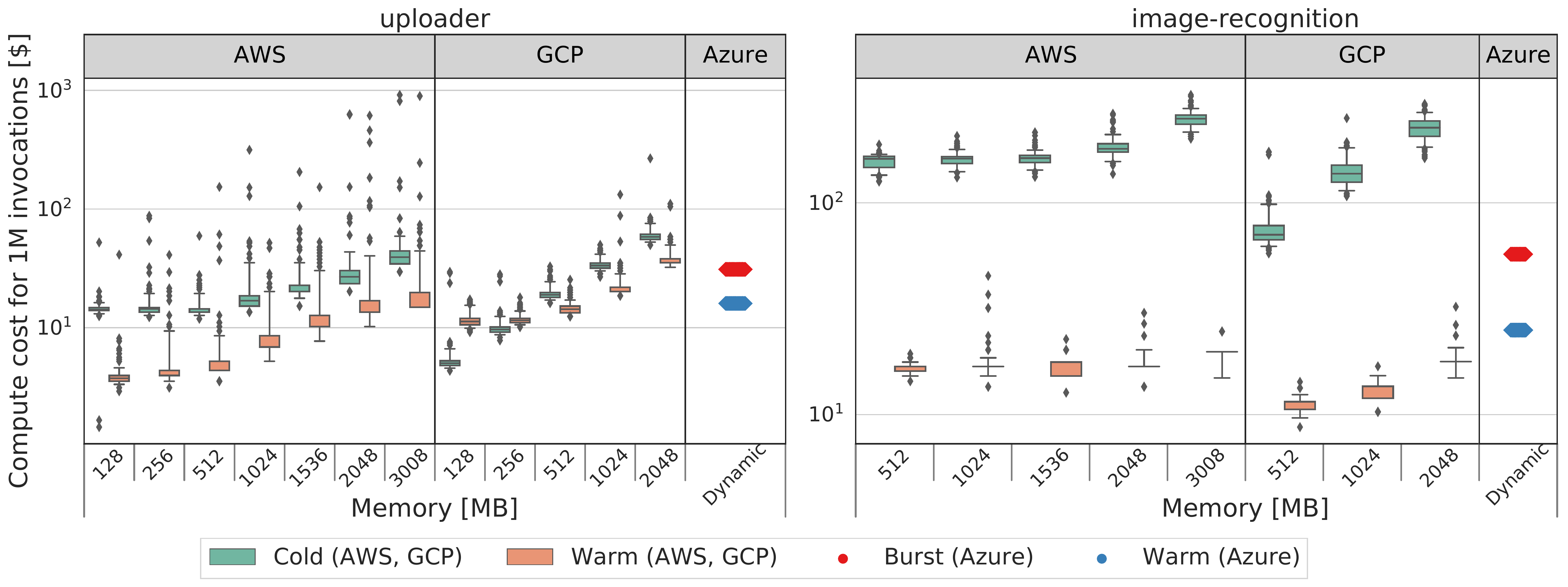}
      \label{fig:cost_compute_time}
    }
	}
	\resizebox*{0.95\width}{0.92\totalheight}{
		\subfloat[Median ratio of used and billed resources (\%).
		]{%
      \includegraphics[width=\dimexpr0.34\textwidth,height=5cm]{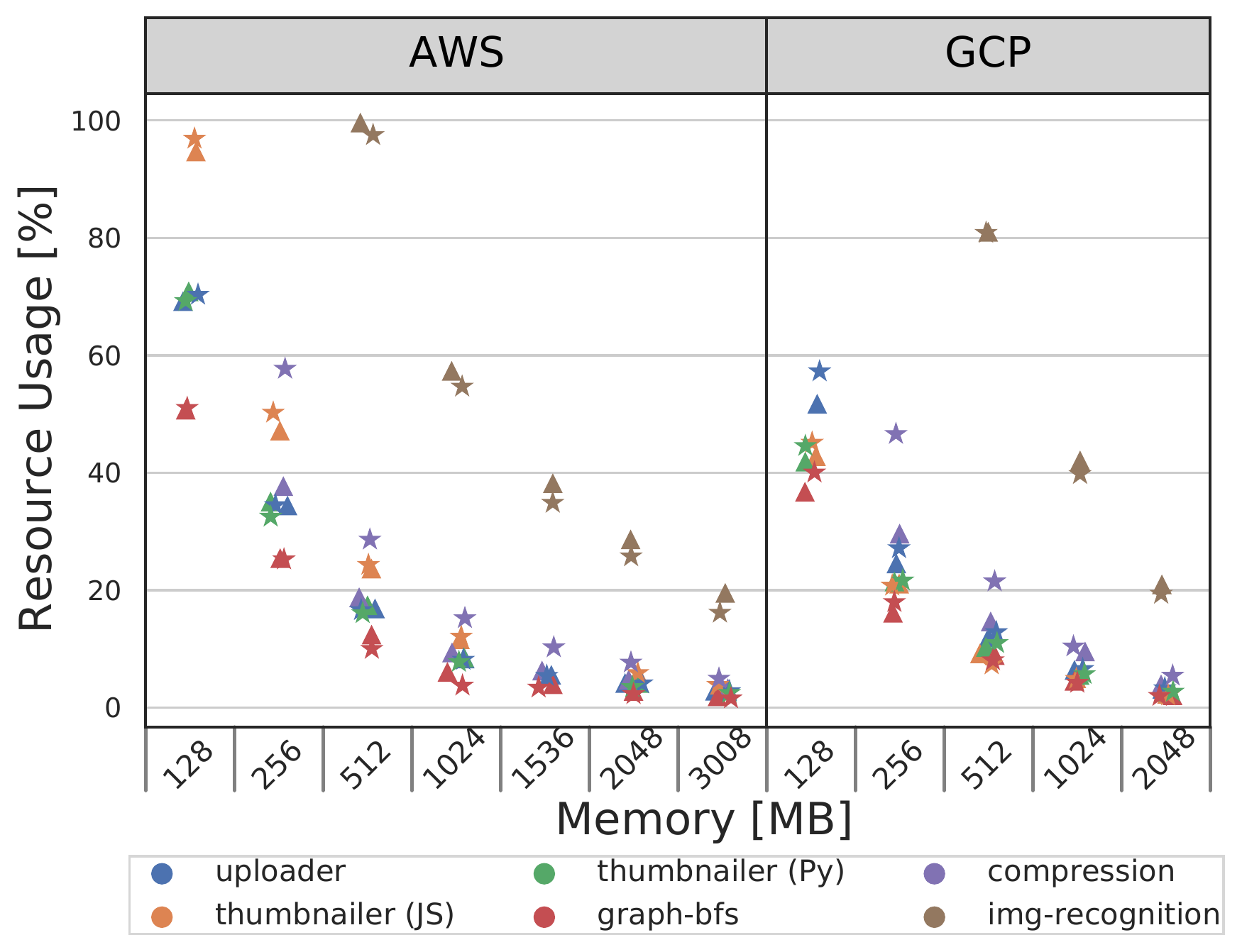}
      \label{fig:resource_usage}
		}
	}
\ifcnf
  \vspace{-1em}
\fi
	\caption{\textbf{
		The cost analysis of performance results from Figure~\ref{fig:time_evaluation}:
    execution cost of 1 million requests (a)
    and resource usage of cold (\ding{115}) and warm (\ding{72}) executions (b).
    Azure data is (a) limited to a single average and (b) not available
    due to limitations of the Azure Monitor systems.
	}}
\end{figure*}

%
%
\emph{\textbf{Q4 FaaS vs IaaS: is serverless slower?}}
The execution environment of serverless functions brings new sources of overheads~\cite{10.1145/3352460.3358296}.
To understand
their impact, we compare serverless performance with their natural alternative: virtual machines,
where the durable allocation and higher price provide a more stable environment and data locality.
We rent an AWS t2.micro instance with one virtual CPU and 1 GB memory
since such instance should have comparable resources with Lambda functions.
We deploy SeBS with the local Docker-based execution environment and measure warm execution times
of 200 repetitions to estimate latency of constantly warm service.
Also, we perform the same experiment with AWS S3 as persistent storage.
This provides a more balanced comparison of performance overheads, as cloud provider storage is commonly used instead of a self-deployed
storage solution, thanks to its reliability and data durability.
We compare the performance against warm provider times (Section~\mbox{\ref{sec:performance}}.Q1),
selecting configurations where benchmarks obtain high performance and further memory increases don't
bring noticeable improvements. We present the summary in Table~\mbox{\ref{tab:iaas_performance}}.
The overheads of FaaS-ifying the service
vary between slightly more than 50\% and a slowdown by a factor of four. Equalizing storage
access latencies reduces the overheads significantly (Python benchmark \emph{thumbnailer}).

\emph{Conclusions: performance overheads of FaaS executions are not uniformly distributed across
application classes. The transition from a VM-based deployment to serverless architecture
will be accompanied by significant performance losses.}

\ifcnf
\vspace{-1em}
\fi
\subsection{Cost Analysis}
\label{sec:cost}

While raw performance may provide valuable insights,
the more important question for systems designers is how much does such performance costs.
We analyze the cost-effectiveness of results from the \emph{Perf-Cost} experiment described, answering four major research questions.

%
\emph{\textbf{Q1 How users can optimize the cost of serverless applications?} }
Each provider includes two major fees in the pay-as-you-go billing model:
a flat fee for 1 million executions and the cost of consumed compute time and memory,
but the implementations are different.
AWS charges for reserved memory and computing time rounded up to 100 milliseconds,
and GCP has a similar pricing model.
On the other hand, Azure allocates memory dynamically and charges for average memory size rounded up to 128 MB.

Since computing and I/O resources are correlated with the amount of requested memory,
increasing memory allocation might decrease execution time and a more expensive memory
allocation might doesn't necessarily lead to an increase in cost.
We study the price of 1 million executions for the I/O-bound \emph{uploader} and
  compute-bound \emph{image-recognition} benchmarks (Figure~\ref{fig:cost_compute_time}).
Performance gains are significant for \emph{image-recognition},
where the cost increases negligibly, but the decreased execution time of \emph{compression} does
not compensate for growing memory costs.
For other benchmarks, we notice a clear cost increase with every expansion of allocated memory.
The dynamic allocation on Azure Functions generates higher
costs, and they cannot be optimized.
\emph{Conclusions: to increase the serverless price-efficiency,
  the user has to not only characterize the requirements
  of their application, but the exact performance boundaries - compute, memory,
I/O - must be learned as well.
Azure Functions generate higher costs because of the dynamic memory allocations.
}

%
\emph{\textbf{Q2 Is the pricing model efficient and fair?} }
FaaS platforms round up execution times and memory consumption, usually to nearest 100 milliseconds
and 128 megabytes.
Thus, users might be charged for unused duration and resources.
With SeBS, we estimate the scale of this problem by comparing actual and billed resource usage.
We use the memory consumption of each invocation and the median memory allocation across the experiment on AWS and GCP, respectively.
We do not estimate efficiency on Azure
because monitor logs contain incorrect information on the memory used\footnote{The issues have been reported to Azure
team.}.

The results in Figure~\ref{fig:resource_usage} show that the required computing power
and I/O bandwidth are not always proportional to memory consumption.
Changing the current system
would be beneficial to both the user and the provider, who could increase the utilization of
servers if declared memory configuration would be closer to actual allocations.
Furthermore, rounding up of execution time affects mostly short-running functions, which
have gained significant traction as simple processing tools for database and messaging
queue events.

\emph{
Conclusions: memory usage is not necessarily correlated with an allocation
of CPU and I/O resources. The current pricing model encourages over-allocation of memory,
leading in the end to underutilization of cloud resources.}

\begin{table}\centering
	\begin{adjustbox}{max width=\linewidth}
  \begin{tabular}{cc|lllllll}\toprule
    &&& \textbf{Upl} & \textbf{Th, Py} & \textbf{Th, JS} & \textbf{Comp} & \textbf{Img-Rec} & \textbf{BFS} \\
    \hline
    \multirow{2}{*}{\rotatebox[origin=c]{90}{IaaS}} & Local & Request/h & 16627 & 79282 & 21697 & 4452 & 17658 & 119272 \\
    & Cloud & Request/h & 11371 & 27503 & 18819 & 1284 & 15312 & 117153 \\
    \hline
    \multicolumn{2}{c|}{\multirow{4}{*}{FaaS}} & Eco 1M [\$] & 3.54 & 2.29 & 3.75 & 32.1 & 15.8 & 2.08 \\
      && Eco B-E & 3275 & 5062 & 3093 & 362 & 733 & 5568 \\
      && Perf 1M [\$] & 6.67 & 3.34 & 10 & 50 & 19.58 & 2.5 \\
      && Perf B-E & 1740 & 3480 & 1160 & 232 & 592 & 4640 \\
		\bottomrule
	\end{tabular}
	\end{adjustbox}
  \caption{\textbf{The break-even point (requests per hour) for the most efficient (Eco) and
      best performing (Perf) AWS Lambda configuration, compared to IaaS deployment (Table~\mbox{\ref{tab:iaas_performance}}).
      IaaS assumes 100\% utilization of the micro.t2 machine costing \$0.0116 per hour.}}
	\label{tab:iaas_cost}
\ifcnf
  \vspace{-2.5em}
\fi
\end{table}

%
\emph{\textbf{Q3 FaaS vs IaaS: when is serverless more cost efficient?}}
The most important advantage of serverless functions is the pay-as-you-go model that enables efficient
deployment of services handling infrequent workloads.
The question arises immediately: how infrequent must be the use of service to achieve lower
cost than a dedicated solution with virtual machines?
The answer is not immediate since the FaaS environment negatively affects the performance (Section~\mbox{\ref{sec:performance}}.Q4).
Thus, we attempt the break-even analysis to determine the maximum workload a serverless function can
handle in an hour without incurring charges higher than a rental.
We summarize in Table~\mbox{\ref{tab:iaas_cost}} the results for the most cost-efficient and the
highest performing deployments of our benchmarks on AWS Lambda.
While EC2-based solution seems to be a clear cost winner for frequent invocations, its scalability
is limited by currently allocated resources.
Adding more machines takes time, and multi-core machines introduce additional cost overheads due to underutilization.
Serverless functions can scale rapidly and achieve much higher throughput.

\emph{
Conclusions: the IaaS solution delivers better performance at a lower price,
but only if a high utilization is achieved.
}

\emph{\textbf{Q4 Does the cost differ between providers?} }
Cost estimations of serverless deployments are usually focused on execution time and
allocated memory, where fees are quite similar (Section~\mbox{\ref{tab:aws_azure}}, Figure~\ref{fig:cost_compute_time}).
There are, however, other charges associated with using serverless functions.
While storage and logging systems are not strictly required, functions must use the provider's
API endpoints to communicate with the outside world.
AWS charges a flat fee for an HTTP API but meters each invocation in 512 kB
increments~\mbox{\cite{awsAPIBiling}}.
GCP and Azure functions are charged \$0.12 and from \$0.04 to \$0.12, respectively,
for each gigabyte of data transferred out~\cite{azureNetworkPricing,gcpPricing}.

Our benchmark suite includes use cases where sending results directly back to the user is the most
efficient way, such as \emph{graph-bfs} returning graph-related data (ca. 78 kB) and
\emph{thumbnailer} sending back a processed image (ca. 3 kB).
The additional costs for one million
invocations can vary from \$1 on AWS to almost \$9 on Google Cloud and Azure\footnote{
HTTP APIs have been available for Lambda since December 2019. REST APIs have higher
fees of \$3.5 for 1M requests and \$0.09 per GB of traffic.
}.

\emph{
Conclusions: billing models of cloud providers include additional charges, and serverless
applications communicating a non-negligible amount of data are particularly affected there.
}

\begin{figure}
    \includegraphics[width=\linewidth]{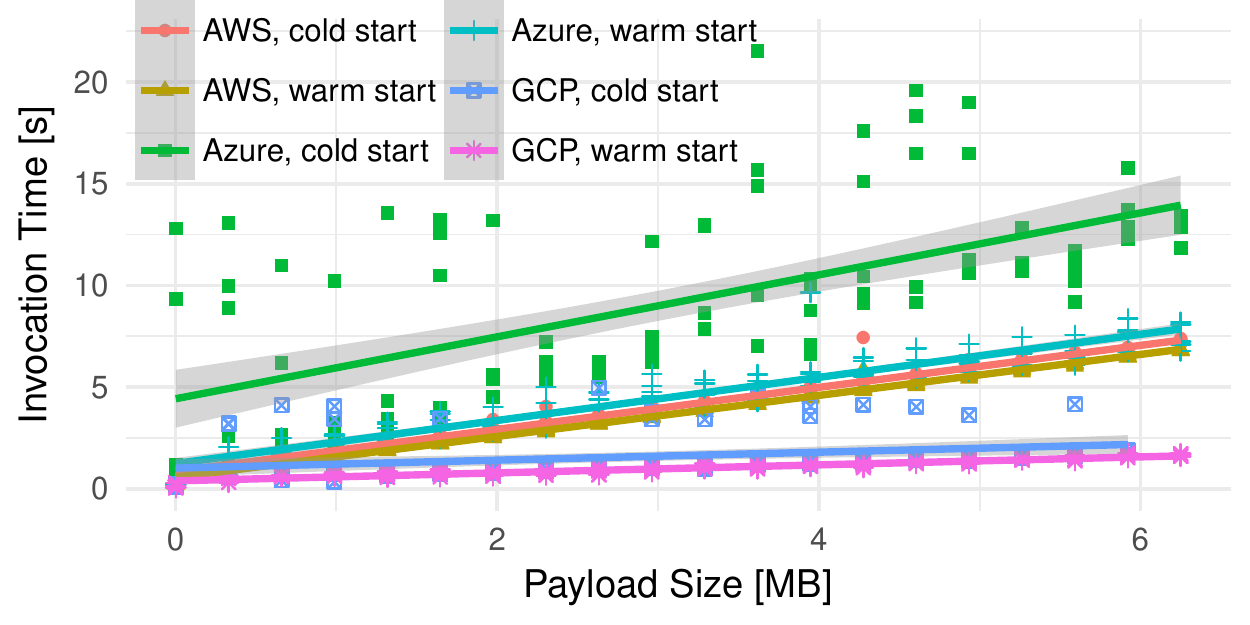}
\ifcnf
    \vspace{-2.5em}
\fi
    \caption{\textbf{Invocation overhead of functions 
    with varying payload.}}
    \label{fig:invoc_overhead}
\ifcnf
    \vspace{-0.5em}
\fi
\end{figure}

\subsection{Invocation overhead analysis }
\label{sec:invocation}

While benchmarking FaaS by saturating bandwidth or compute power tells
a part of a story, these services are not designed with such workloads in mind. 
On the contrary, they are intended to handle large amounts of 
smaller requests, often arriving in unpredictable behavior,
where the latency of starting the system may play a pivotal 
role.
The function invocation latency
depends on factors that are hidden from the user (Section~\ref{sec:platform_model}),
and performance results indicated that FaaS systems add non-trivial overheads
there (Section~\ref{sec:performance}).

As there are no provider-specific APIs to query such metrics,
users must estimate these overheads by comparing client-side round-trip latency
and function execution time.
However, such comparison is meaningful only for symmetric connections.
That assumption doesn't hold for serverless:
invocation includes the overheads of FaaS controllers, whereas returning results
should depend only on network transmission.
Instead, we estimate latencies of black-box invocation systems accurately
with a different approach in the experiment \textbf{Invoc-Overhead}.
First, we use timestamps to measure the time that passes between invocation
and the execution start, considering all steps, including
language worker process overheads. To compare timestamps, we follow an existing
clock drift estimation protocol~\cite{4536494}.
%
We measure the invocation latency in regions \emph{us-east-1}, \emph{eastus},
and \emph{us-east1} on AWS, Azure, and GCP, respectively.
We analyze round-trip times and discover that
they follow an asymmetric distribution, as in~\cite{4536494}. Thus, for clock
synchronization, we exchange messages until seeing no lower round-trip time in
$N$ consecutive iterations. We pick $N = 10$, since the relative difference between
the lowest observable connection time and the minimum time after ten
non-decreasing connection times is ca.~5\%.
Using the benchmarking methodology outlined above, we analyze how the invocation overhead
depends on the function input size for 1kB--5.9MB (6MB is the limit for AWS endpoints).
The results presented in~\autoref{fig:invoc_overhead} show the latency cost of
cold and warm invocations.

\emph{\textbf{Q1 Is the invocation overhead consistent?} }
We found that invocation latency behavior to be fairly consistent and predictable for cold AWS
runs and warm startups on both platforms. 
At the same time, cold startups on Azure and GCP cannot be easily explained.
Similarly to findings in Section~\mbox{\ref{sec:performance}}, we observe a cold start
behavior that can be caused by unpredictable delays when scheduling functions in the cloud,
or the overheads associated with an inefficient implementation of local servers executing the function.
\emph{Conclusions: while warm latencies are consistent and predictable,
  cold startups add unpredictable performance deviations into serverless applications on Azure and GCP.
}

%
\emph{\textbf{Q2 Does the latency change linearly with an increase in payload size?} }
With the exception of Azure's and GCP's cold starts, the latency scales linearly. For warm invocations
on AWS, Azure, and GCP, and cold executions on AWS, the linear model fits almost
perfectly the measured data, with adjusted R\textasciicircum2 metric 0.99, 0.89, 0.9, and 0.94,
respectively. 
\emph{Conclusions: network transmission times is the only major overhead associated
  with using large function inputs.
}

%
%


\ifcnf
\begin{figure*}[t]
\vspace{-1.25em}
	\centering
	\subfloat[\footnotesize\rm\textbf{Language: NodeJs, memory 
	allocated: 128 MB, function execution time: 1s.}]
	{\includegraphics[width=\fw \textwidth]
		{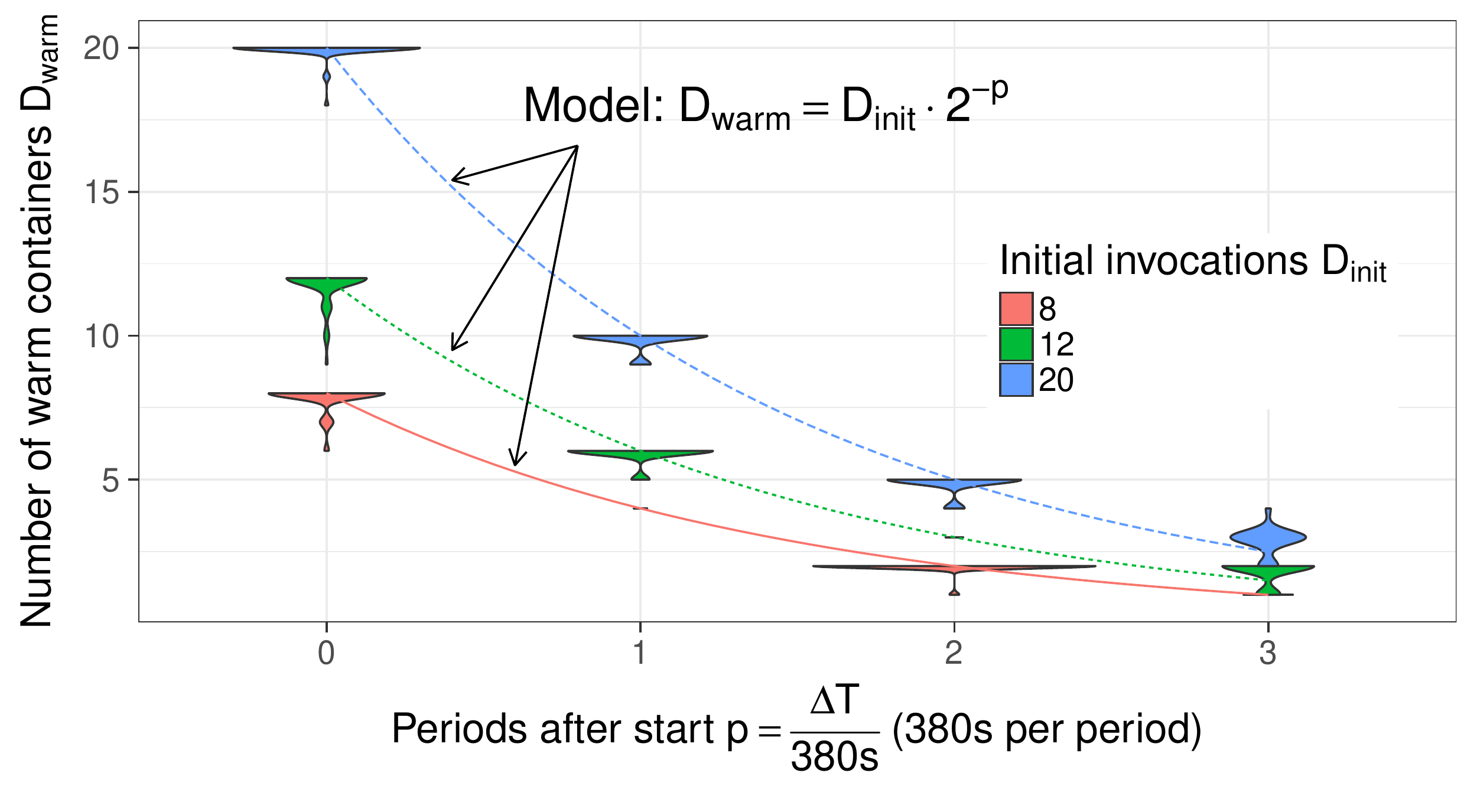}
		\label{fig:nodejs_128_1}}
	%
%
	\hfill
	\subfloat[\footnotesize\rm\textbf{Language: Python, memory 
	allocated: 1536 MB, function execution time: 1s.}]
	{\includegraphics[width=\fw \textwidth]			
		{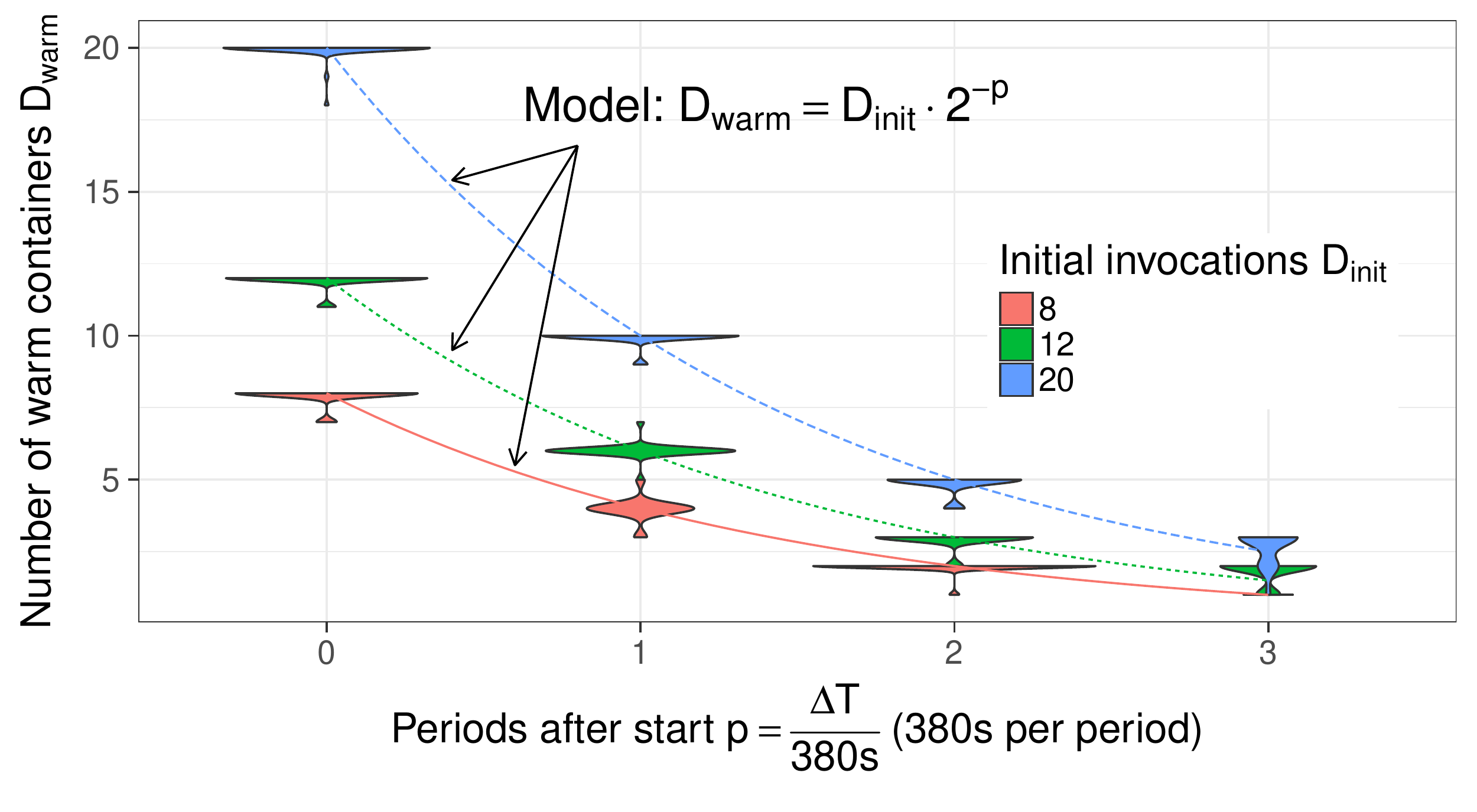}
		\label{fig:python_1536_1}}
	\hfill
	\subfloat[\footnotesize\rm\textbf{Language: Python, 
	memory 
		allocated: 1536 MB, function execution time: 10s.}]
	{\includegraphics[width=\fw \textwidth]			
		{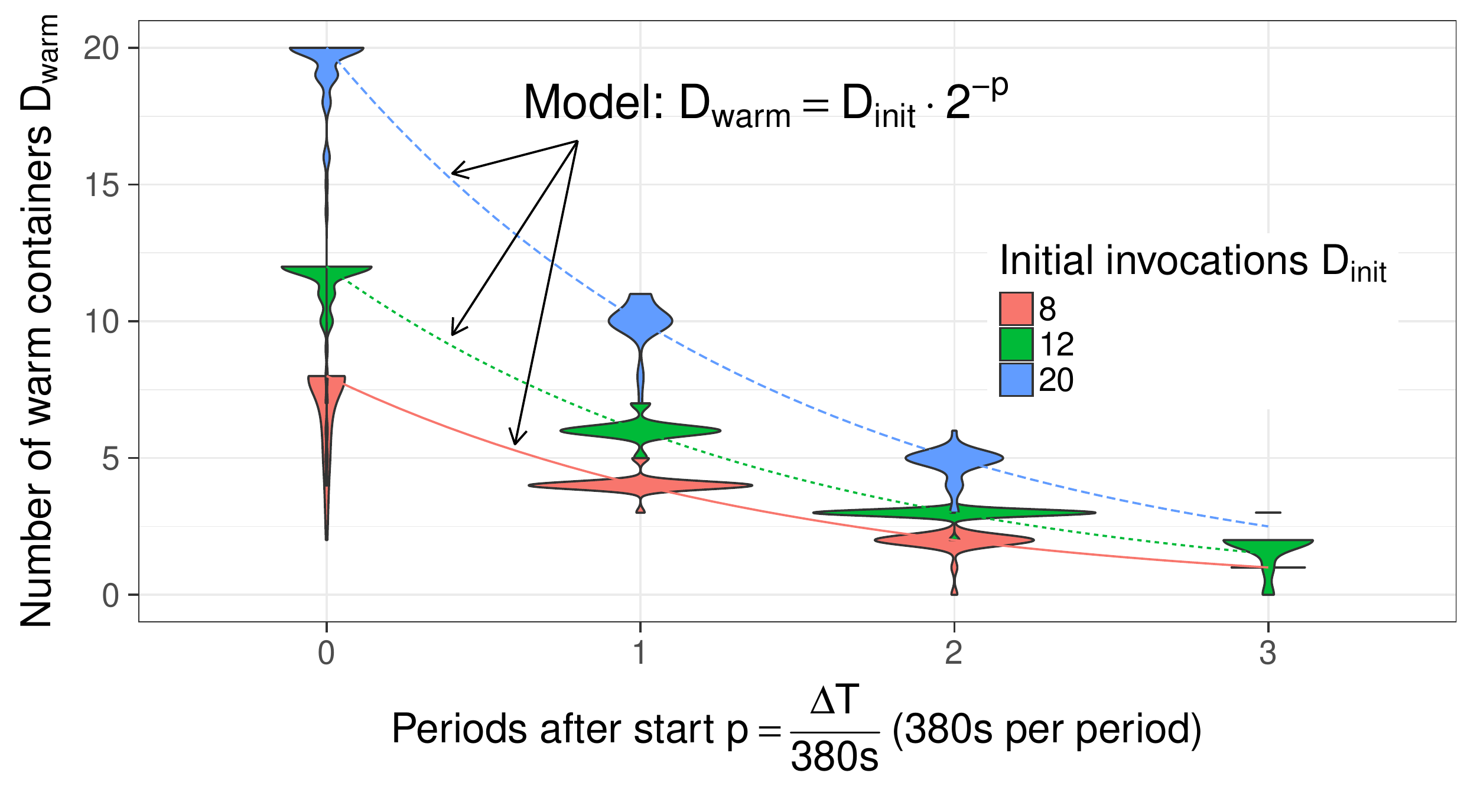}
		\label{fig:python_1536_10}}	
	%

%
%
%

	\vspace{-1em}
	\caption{
		\vspace{-1em}
		\textbf{{Representative scenarios of eviction 
		policies of FaaS containers on AWS.
	} }
		\vspace{-0.5em}
	}
	\label{fig:performancePlotsSquare}
	
\end{figure*}
\else
  \begin{figure*}[t]
    \centering
    \subfloat[Language: NodeJs, memory allocated: 128 MB, function execution time: 1s.]
    {\includegraphics[width=\fw \textwidth]
      {figures/eviction-model/aws_nodejs_sleep_1_results_128_1}
      \label{fig:nodejs_128_1}}
  	\hfill
  	\subfloat[Language: Python, memory allocated: 128 MB, function execution time: 1s.]
  	{\includegraphics[width=\fw \textwidth]			
  	{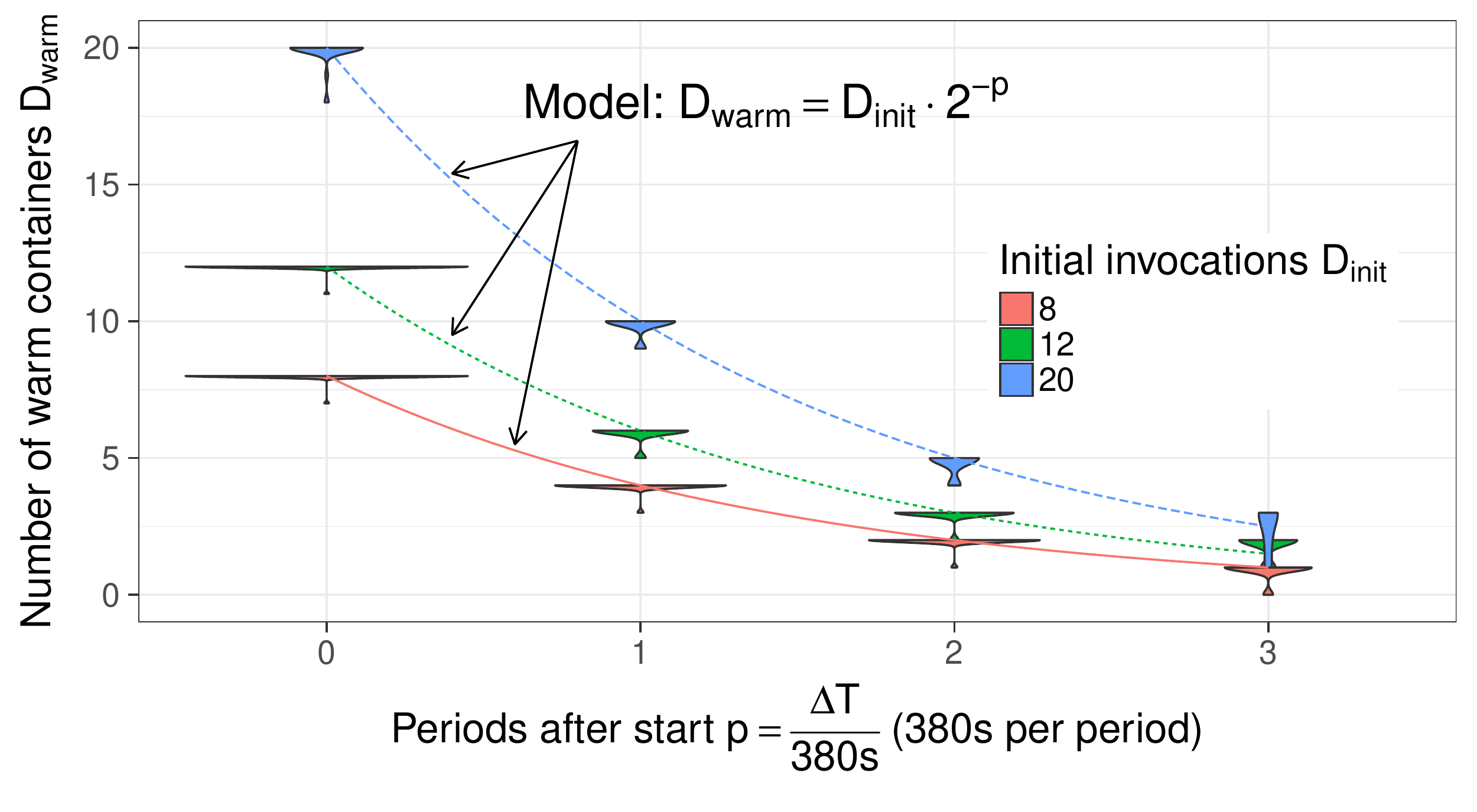}
  		\label{fig:python_128_1}}	
    \hfill
    \subfloat[Language: Python, memory allocated: 1536 MB, function execution time: 1s.]
    {\includegraphics[width=\fw 
      \textwidth]			
      {figures/eviction-model/aws_python_sleep_1_results_1536_1}
      \label{fig:python_1536_1}}

  	\subfloat[Language: Python, memory allocated: 128 MB, function execution time: 10s.]
    {
      \includegraphics[width=\fw \textwidth]{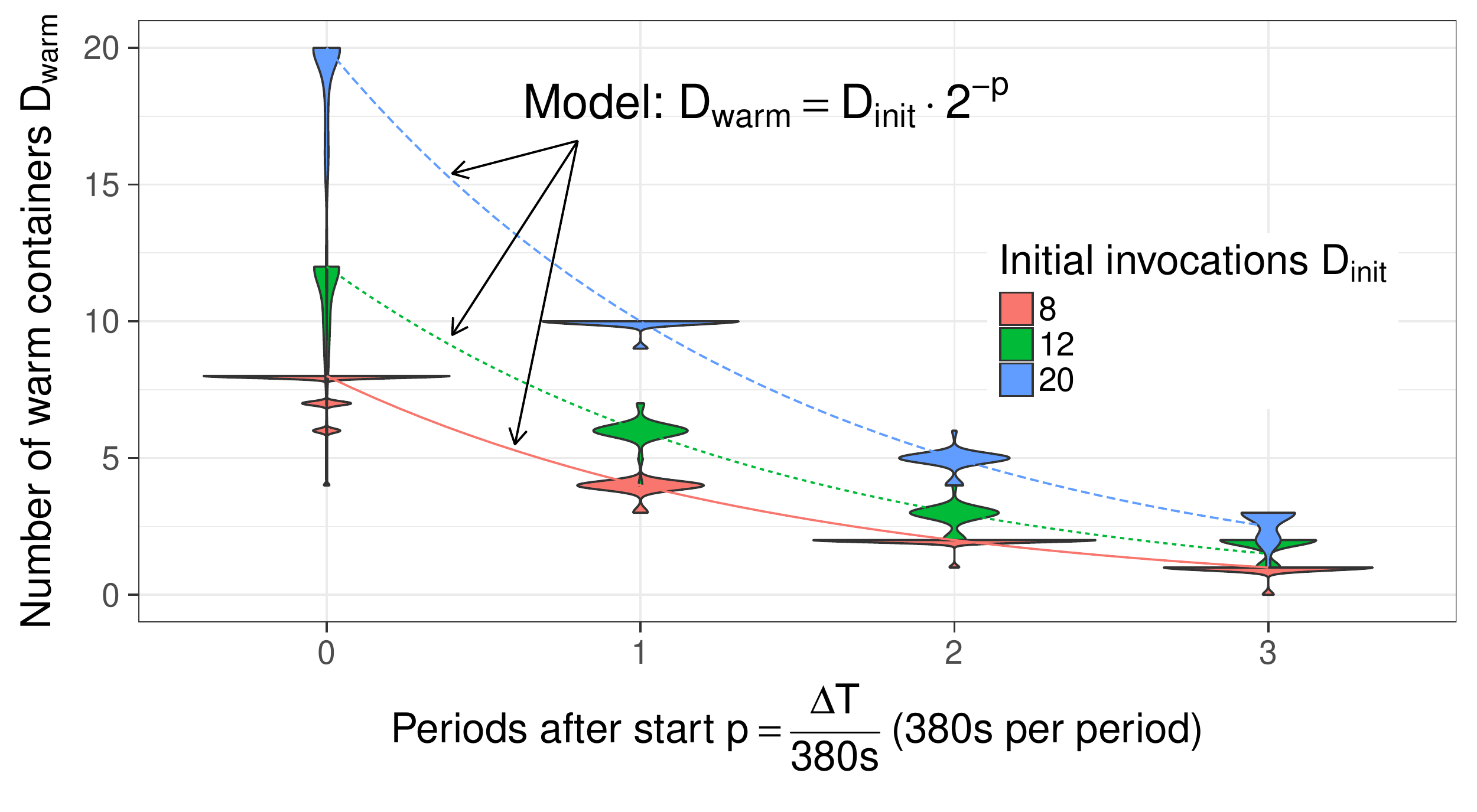}
      \label{fig:nodejs_128_10}
    }
  \hfill
    \subfloat[Language: Python, memory allocated: 1536 MB, function execution time: 10s.]
    {
      \includegraphics[width=\fw \textwidth]{figures/eviction-model/aws_python_sleep_10_results_1536_10}
      \label{fig:python_1536_10}
    }
  \hfill
  \subfloat[Language: Python, memory alloc.: 128 MB, function exec.~time: 1s. Code package 250 MB.]
    {
      \includegraphics[width=\fw \textwidth]{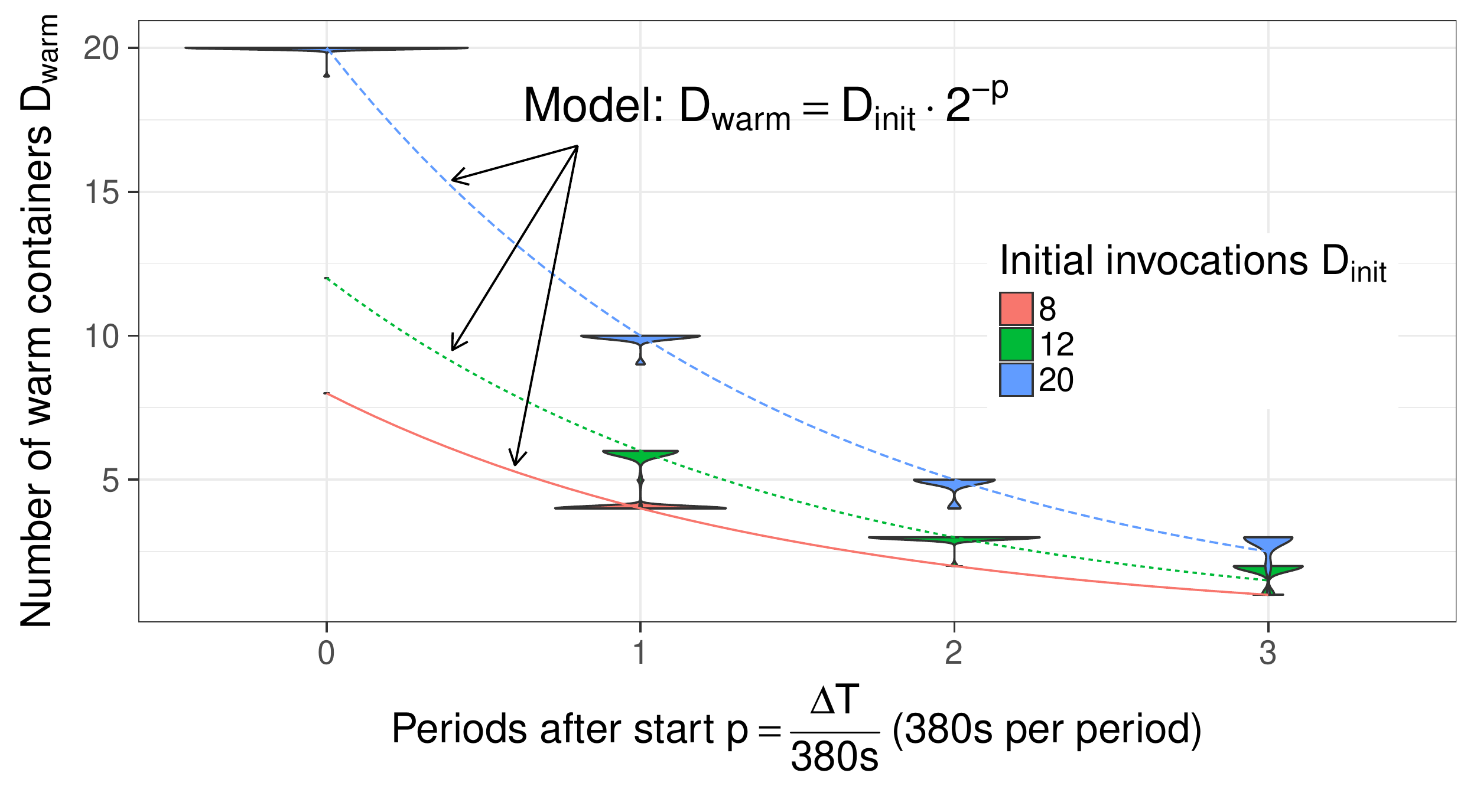}
      \label{fig:python_128_1_heavy}
    }

    \caption{
      \textbf{{Representative scenarios of eviction 
      policies of FaaS containers on AWS.
    } }
    }
    \label{fig:performancePlotsSquare}
    
  \end{figure*}
\fi

\begin{table}\centering
\ifcnf
\footnotesize
\fi
\begin{adjustbox}{max width=\linewidth}
		\begin{tabular}{llll}\toprule
		\textbf{Parameter} & \textbf{Range} & \textbf{Parameter} & \textbf{Range} \\
		\midrule
		$D_{init}$ & 1-20 & $\Delta T$ & 1-1600 s \\
		Memory & 128-1536 MB & Sleep time & 1-10 s\\
		Code size & 8 kB, 250 MB & Language & Python, Node.js \\
		\bottomrule
	\end{tabular}
\end{adjustbox}
	\caption{\textbf{Container eviction experiment parameters.}}
	\label{tab:model_params}
\ifcnf
  \vspace{-3.5em}
\fi
\end{table}

%
\subsection{Container eviction model. }
\label{sec:eviction}

In the previous section, we observed a significant 
difference in startup 
times depending on whether we hit a cold or warm start. 
Now we analyze how we can increase the chance of hitting 
warm containers by adjusting the invocation policy. 
Yet, service providers do not publish their policies. Thus, to guide users, we
created the experiment \textbf{Eviction-Model} to empirically model function's cold
start characteristics.

\emph{\textbf{Q1 Are cold starts deterministic, repeatable, and application agnostic?} }
The container eviction policy can depend on function
parameters like number of previous invocations, allocated memory size,
execution time, and on global factors: system occupancy or hour.
Hence, we use the following benchmarking setup: at a particular time,
we submit $D_{init}$ initial invocations, we wait $\Delta T$ seconds, and then
check how many $D_{warm}$ containers are still active. Next, we test
various combinations of $D_{init}$ and $\Delta T$ for different
function properties (Table~\ref{tab:model_params}).
%
Our results reveal that the \textbf{AWS} container eviction policy
is surprisingly agnostic to many function properties: allocated memory, execution time,
language, and code package size.
Specifically, after every 380
seconds, half of the existing containers are evicted. The container lifecycles 
are shown in Figures~\ref{fig:nodejs_128_1}-
\ifcnf
  \ref{fig:python_1536_10}. 
\else
\ref{fig:python_128_1_heavy}. 
\fi

We also attempted to execute these benchmarks on \textbf{Azure} Functions. Yet,
massive concurrent invocations led to random and unpredictable failures when
invoking functions.
\emph{Conclusions: the eviction policy of containers is deterministic and
  application agnostic, and cold startup frequency can be predicted when scaling serverless applications.}

\emph{\textbf{Q2 Can we provide simple analytical eviction models?} }
In Equation~\ref{eq:aws_model}, we provide a simple analytical model of the number of active containers.
The model fits the data (Figures~\ref{fig:nodejs_128_1}-\ref{fig:python_1536_10}) extremely well.
\ifcnf
\vspace{-0.4em}
\fi
\begin{equation}
\label{eq:aws_model}
D_{warm} = D_{init} \cdot 2^{-p},\quad p = \left\lfloor {\Delta T}/{380s} \right 
\rfloor
\end{equation}
\ifcnf
\vspace{-0.2em}
\fi
We performed a well-established $R^2$ statistical test to validate its
correctness, and the $R^2$ metric is more than 0.99. The only exception are
Python experiments with 10s sleep time, yet even there $R^2$ is
$>$0.94.
Thus, we can use Model~\ref{eq:aws_model} to find the time-optimal invocation
batch size $D_{init}$, given that user needs to run $n$ instances of a function
with runtime $t$:
\ifcnf
\vspace{-0.4em}
\fi
\begin{equation}
\label{eq:aws_sol}
	D_{init, opt} = {n\cdot t}/{P}
\end{equation}
\ifcnf
\vspace{-0.2em}
\fi
\noindent
where $P = 380s$ is the AWS eviction period length.

\emph{Conclusions: we derive an analytical model for cold start frequency that can
  be incorporated into an application to warm up containers and avoid cold starts,
  without using provisioned concurrency solutions that have a non-serverless billing model.
}

%% file: secs/related-work.tex
\input{secs/table-rw}

\begin{table}\centering
\ifcnf
  \vspace{-1em}
\fi
\setlength{\tabcolsep}{1pt}
	\begin{adjustbox}{max width=\linewidth}
    \begin{tabular}{ll}\toprule
      \makecell[l]{\textbf{Results, methods, and insights}} & \makecell[c]{\textbf{Novel}\\ \textbf{insights?}} \\
		\midrule
    AWS Lambda achieves the best performance on all workloads. & \faTimes \cite{Maissen_2020,9027346} \\
    Irregular performance of concurrent Azure Function executions. & \faTimes\ \cite{Maissen_2020} \\
    I/O-bound functions experience very high latency variations. & \faTimes\ \cite{10.5555/3277355.3277369} \\
    High-memory allocations increase cold startup overheads on GCP. & \faThumbsUp \\
    GCP functions experience reliability and availability issues. & \faThumbsUp \\
    \makecell[l]{AWS Lambda performance is not competitive against VMs\\ assuming comparable resources.} & \faThumbsUp \\
		\midrule
    High costs of Azure Functions due to unconfigurable deployment.   & \faThumbsUp \\
    Resource underutilization due to high granularity of pricing models. & \faThumbsUp \\
    \textbf{Break-even analysis for IaaS and FaaS deployment.} & \faTimes\ \cite{Mller2019LambadaID} \\
    The function output size can be a dominating factor in pricing. & \faThumbsUp \\
		\midrule
    \textbf{Accurate methodology for estimation of invocation latency.} & \faThumbsUp \\
    Warm latencies are consistent and depend linearly on payload size. & \faThumbsUp \\
    Highly variable and unpredictable cold latencies on Azure and GCP. & \faTimes\ \cite{Maissen_2020} \\
		\midrule
    AWS Lambda container eviction is agnostic to function properties. & \faTimes\ \cite{10.5555/3277355.3277369} \\
    \textbf{Analytical models of AWS Lambda container eviction policy.} & \faTimes\ \cite{10.5555/3277355.3277369} \\
		\bottomrule
	\end{tabular}
\end{adjustbox}
\caption{\textbf{The impact of SeBS:} insights and new methods (\textbf{bolded}) provided in our work,
  with a comparison against similar results in prior work.}
	\label{tab:rw_table_insights}
\ifcnf
  \vspace{-2em}
\fi
\end{table}

%
We summarize contributions to serverless benchmarking
in~\autoref{tab:comparison}.
We omit the microservices benchmark suite by Gan et al.~\cite{Gan2019} as it
includes only one FaaS benchmark evaluation.
Contrary to many existing cloud benchmarks, SeBS offers a systematic approach
where a diverse set of real-world FaaS applications is used instead of
microbenchmarks expressing only basic CPU, memory, and I/O requirements.
While newer benchmark suites started to adopt automatic deployment and
invocation, the installation of dependencies in a compatible way is often
omitted when only microbenchmarks are considered. With a vendor-independent
benchmark definition and focus on binary compatibility, we enable
the straightforward inclusion of new benchmarks representing emerging solutions and
usage patterns.

SeBS brings new results and insights (Table~\ref{tab:rw_table_insights}).
SeBS goes beyond just an evaluation of the function latency and throughput~\cite{Back2018,Kim2019},
and we focus on FaaS consistency, efficiency, initialization overheads,
and container eviction probabilities
and include a set of local metrics that are necessary to characterize resource consumption accurately.
Only a single work takes HTTPS overheads
into account when measuring invocation latencies~\cite{faasbenchmark}.
Our work confirms findings that AWS Lambda provides overall the best performance~\cite{Maissen_2020,9027346},
but we obtain this result with a diverse set of practical workloads instead of microbenchmarks.
We generalize and extend past results on container eviction modeling~\cite{10.5555/3277355.3277369}
and the break-even analysis for a specific application~\cite{Mller2019LambadaID}.
Maissen et al.~\cite{Maissen_2020} report several similar findings to ours.
However, their invocation latency estimations depend on round-trip measurements,
and they don't take payload size into account.
Similarly to us, Yu et al.~\cite{10.1145/3419111.3421280} find other sources of cost inefficiency on AWS.
However, their benchmark suite is focused on function
composition and white-box, open-source platforms OpenWhisk and Fn.
We provide insights into the black-box
commercial cloud systems to understand the
reliability, and economics of the serverless middleware.

\sloppy
Finally, there are benchmarks in other domains: SPEC~\cite{spec} (clouds, HPC)
LDBC benchmarks (graph analytics)~\cite{iosup2016ldbc},
GAPBS~\cite{beamer2015gap} and Graph500~\cite{murphy2010introducing} (graph related),
DaCapo~\cite{10.1145/1167515.1167488} (Java applications),
Deep500~\cite{ben2019modular} and MLPerf~\cite{reddi2019mlperf} (machine learning), and
Top500~\cite{dongarra1997top500} (dense linear algebra).

%% file: secs/table-rw.tex
\begin{table}[t]
%
\setlength{\tabcolsep}{0.5pt}
\renewcommand{\arraystretch}{0.7}
\centering
\footnotesize
\scriptsize
%
\begin{tabular}{lccccccccccccccccc}
\toprule
\multirow{2}{*}{\makecell[c]{\textbf{Reference, }\\ \textbf{Infrastructure}}} & 
\multicolumn{6}{c}{\makecell[c]{\textbf{Workloads}}} & 
\multicolumn{3}{c}{\makecell[c]{\textbf{Lang.}}} & 
\multicolumn{4}{c}{\makecell[c]{\textbf{Platform}}} & 
\multicolumn{4}{c}{\makecell[c]{\textbf{Infrastructure}}} \\
\cmidrule(lr){2-7} \cmidrule(lr){8-10} \cmidrule(lr){11-14} \cmidrule(lr){15-18}
 & 
\textbf{Mc} & 
\textbf{Wb} & 
\textbf{Ut} & 
\textbf{Ml} & 
\textbf{Sc} & 
\textbf{ML} &
\textbf{Py} &
\textbf{JS} &
\textbf{OT} &
\textbf{AW} &
\textbf{AZ} &
\textbf{GC} &
\textbf{OT} &
%
%
\textbf{Im} &
\textbf{At} &
\textbf{Dp} &
\textbf{Nw} 
 \\
\midrule
FaaSTest~\cite{faasbenchmark} & 
\faThumbsOUp & \faThumbsDown & \faThumbsDown & \faThumbsDown & \faThumbsDown &\faThumbsOUp &
\faThumbsOUp & \faThumbsUp & \faThumbsUp &
\faThumbsOUp & \faThumbsOUp & \faThumbsOUp & \faThumbsDown &
\faThumbsOUp & \faThumbsOUp & \faThumbsDown & \faThumbsDown \\ 
FaasDom~\cite{faasdom,Maissen_2020} & 
\faThumbsOUp & \faThumbsDown & \faThumbsDown & \faThumbsDown & \faThumbsDown &\faThumbsDown &
\faThumbsOUp & \faThumbsOUp & \faThumbsOUp &
\faThumbsOUp & \faThumbsOUp & \faThumbsOUp & \faThumbsOUp &
\faThumbsOUp & \faThumbsOUp & \faThumbsDown & \faThumbsOUp \\ 
Somu et al.~\cite{9027346} & 
\faThumbsOUp & \faThumbsDown & \faThumbsDown & \faThumbsDown & \faThumbsDown &\faThumbsDown &
\faThumbsOUp & \faThumbsDown & \faThumbsDown &
\faThumbsOUp & \faThumbsDown & \faThumbsOUp & \faThumbsDown &
\faThumbsOUp & \faThumbsOUp & \faThumbsDown & \faThumbsDown \\
EdgeBench~\cite{Das2018} & 
\faThumbsDown & \faThumbsDown & \faThumbsOUp & \faThumbsOUp & \faThumbsOUp &\faThumbsOUp  &
\faThumbsOUp & \faThumbsDown & \faThumbsDown &
\faThumbsOUp & \faThumbsOUp & \faThumbsDown & \faThumbsDown &
\faThumbsOUp & \faThumbsDown & \faThumbsDown & \faThumbsDown \\
Kim et al.~\cite{Kim2019, Kim20192} & 
\faThumbsOUp & \faThumbsOUp & \faThumbsOUp & \faThumbsOUp & \faThumbsOUp &\faThumbsOUp  &
\faThumbsOUp & \faThumbsDown & \faThumbsDown &
\faThumbsOUp & \faThumbsOUp & \faThumbsOUp & \faThumbsDown &
\faThumbsOUp & \faThumbsDown & \faThumbsDown & \faThumbsDown \\
Serverlessbench~\cite{10.1145/3419111.3421280} & 
\faThumbsOUp & \faThumbsDown & \faThumbsOUp & \faThumbsOUp & \faThumbsDown &\faThumbsDown &
\faThumbsOUp & \faThumbsOUp & \faThumbsOUp &
\faThumbsOUp & \faThumbsDown & \faThumbsDown & \faThumbsOUp &
\faThumbsOUp & \faThumbsUp & \faThumbsDown & \faThumbsOUp \\
Back et al.~\cite{Back2018} & 
\faThumbsDown & \faThumbsDown & \faThumbsDown & \faThumbsDown & \faThumbsOUp &\faThumbsDown &
\faThumbsDown & \faThumbsOUp & \faThumbsDown &
\faThumbsOUp & \faThumbsOUp & \faThumbsOUp & \faThumbsOUp &
\faThumbsOUp & \faThumbsUp & \faThumbsDown & \faThumbsDown \\
\midrule
\textbf{SeBS [This work]} & 
\faThumbsOUp & \faThumbsOUp & \faThumbsOUp & \faThumbsOUp & \faThumbsOUp & \faThumbsOUp &
\faThumbsOUp & \faThumbsOUp & \faThumbsDown &
\faThumbsOUp & \faThumbsOUp & \faThumbsOUp & \faThumbsDown &
\faThumbsOUp & \faThumbsOUp & \faThumbsOUp & \faThumbsOUp \\
\bottomrule
\end{tabular}
%
%
\caption{\textmd{
\textbf{Related work analysis}: a comparison of existing serverless- and
FaaS-related benchmarks, focusing on supported workloads and functionalities
(we exclude \emph{proposals not followed by the actual
benchmarks}~\cite{cfServerless, vanEyk2018}).
\ul{\textbf{Workloads}}:
\textbf{Mc} (microbenchmarks),
\textbf{Wb} (web applications),
\textbf{Ut} (utilities),
\textbf{Ml} (multimedia),
\textbf{Sc} (scientific),
\textbf{ML} (Machine learning inference).
\ul{\textbf{Languages}}: available implementations, 
\textbf{Py} (Python),
\textbf{JS} (Node.js),
\textbf{OT} (Other),
\ul{\textbf{Platform}}:
\textbf{AW} (AWS Lambda),
\textbf{AZ} (Azure Functions),
\textbf{GC} (Google Cloud Functions),
\textbf{OT} (Other commercial/open-source platforms),
\ul{\textbf{Infrastructure}}:
\textbf{Im} (public implementation),
\textbf{At} (automatic deployment and invocation),
\textbf{Dp} (building dependencies in compatible environment),
\textbf{Nw} (delivered new insights or speedups).
\faThumbsOUp: Supported. \faThumbsUp: Partial support. 
\faThumbsDown: no support.
}}
\ifcnf
\vspace{-1em}
\fi
\label{tab:comparison}
\end{table}

%% file: secs/conclusion.tex
We propose SeBS: benchmark suite
that facilitates developing, evaluating, and analyzing emerging
Function-as-a-Service cloud applications. To design SeBS, we first
develop a simple benchmarking model that abstracts away details of cloud
components, enabling portability. Then, we deliver a specification and
implementation with a wide spectrum of metrics, applications,
and performance characteristics.
%

We evaluated SeBS on the three most popular FaaS providers: AWS, Microsoft
Azure, and Google Cloud Platform.
Our broad set of results show that (1) AWS is considerably faster in
almost all scenarios, (2) Azure suffers from high variance, (3)
performance and behavior are not consistent across providers,
(4) certain workloads are less suited for serverless deployments
and require fine-tuning.
We provide new experimental insights, from performance overheads and portability,
through cost-efficiency, and developed models for container eviction
and invocation overhead. We showed that our open-source benchmark suite gives
users an understanding and characterization of the serverless middleware
necessary to build and optimize applications using FaaS as its execution
backend on the cloud.